\documentclass[%
reprint,
superscriptaddress,
amsmath,amssymb,
aps,
prc,
]{revtex4-1}

\usepackage{graphicx}
\usepackage{dcolumn}
\usepackage{bm}
\newcommand{\Frac}[2]{\frac
{\textstyle\lefteqn{\phantom{{}_{\mathstrut}^{\mathstrut}}} #1}
{\textstyle\lefteqn{\phantom{{}_{\mathstrut}^{\mathstrut}}} #2}}
\newcommand{\efrac}[2]{\frac{\scriptstyle \mathstrut #1}
{\scriptstyle \mathstrut #2}}

\begin{document}

\title{Neutrino recoil force in electron-capture decay of polarized nuclei: measurement prospects and potential applications}
\author{A. L. Barabanov}
\affiliation{National Research Centre ''Kurchatov Institute'', 123182, Moscow, Russia}
\affiliation{Moscow Institute of Physics and Technology, 141701, Dolgoprudny, Moscow Region, Russia}
\author{O. A. Titov}
\email{titov\_oa@nrcki.ru}
\affiliation{National Research Centre ''Kurchatov Institute'', 123182, Moscow, Russia}

\begin{abstract}
Due to a directional asymmetry of neutrino emission caused by parity violation, a sample of radioactive atoms experiences a small recoil force from neutrino radiation accompanying electron capture by polarized nuclei. An expression for this force is obtained for the case of allowed nuclear transitions. Prospects to measure this force by the use of modern micromechanical devices are considered. Numerical estimates for the force are presented for a number of most suitable radioactive isotopes. Potential applications for the weak interaction studies are discussed including the possibility to search for hypothetical Lorentz invariance violation.
\end{abstract}


\maketitle

\section{Introduction}\label{s1}

Discussion on new feasible neutrino sources for neutrino oscillation studies contributed to renewed interest in electron capture (EC) because EC-unstable ions generate monochromatic electron neutrinos. Proposals \cite{Bernabeu2005,Sato2005} for the EC based neutrino sources appeared as a part of a more general concept to use $\beta$-decaying nuclei (or ions with such nuclei) accelerated and accumulated in a storage ring to produce intense collimated neutrino beams ($\beta$ beams)~\cite{Zucchelli2002} (for more details on $\beta$ beam projects, see \cite{Edgecock2013,Wildner2014}). With neutron-deficient nuclei decaying via $\beta^+$ or EC, one can form modulated completely or partially monochromatic neutrino beams ($\beta$ and EC beams); such beams could be useful for neutrino oscillation experiments, as well as for other investigations in physics of the weak interaction (see \cite{Barabanov2015,Barabanov2017} and references therein).

Recent study \cite{DeAngelis2012} (see also \cite{Folan2014,Folan2017}), initiated by proposals \cite{Bernabeu2005,Sato2005,Zucchelli2002} on collimated neutrino beams, is devoted to another aspect of EC. It concerns a recoil force resulted from the asymmetry of neutrino emission in EC involving polarized nuclei. To begin, let a nucleus with spin~$J_i$ capture an electron from the $x$-shell, undergo a transition to the $n$-th state of the daughter nucleus with spin~$J_f$ and energy $E_n^*$ ($E_0^*=0$ for transitions to the ground state $n=0$) and emit an electron neutrino with energy
\begin{equation}\label{1}
E_{\nu nx}=Q_{nEC}-E_x-E_R,\quad Q_{nEC}=\Delta(mc^2)-E_n^*.
\end{equation}
Here $\Delta(mc^2)$ is the difference of the rest energies of the initial and final atoms, $E_x$ is the final atom excitation energy (the atom has a hole in the $x$-shell), $E_R\simeq E_{\nu nx}^2/(2m_fc^2)$ is the recoil energy for the final atom of mass $m_f$.

The simple relation between $E_R$ and $E_{\nu nx}$ is due to the fact that the final state consists of only two particles, neutrino and final atom, so the energies of these particles are fixed by energy and momentum conservation laws. The recoil energy $E_R$ is very small and is usually neglected. However, in the early studies of weak interactions the recoil effect in elementary processes was a subject of careful analysis as an indirect evidence for the neutrino existance. The first successful measurement of $E_R$ was reported for the reaction $^7$Be$+e^-\to\, ^7$Li$+\nu_e$~\cite{Allen1942}.

If there is no preferred direction in space, then the angular distribution of neutrinos (and recoil atoms) is isotropic. In particular, in the experiment \cite{Allen1942}, radioactive $^7$Be atoms, deposited on a platinum foil, received a recoil momentum to the foil or in the opposite direction with equal probability. In the latter case, the ions escaped the metal surface and reached spectrometer that measured their kinetic energy. However, when the nuclei are polarized along some axis $z$, the angular distribution is anisotropic due to parity violation. Then, if, for example, the neutrinos are emitted predominantly opposite to $z$-axis, the recoil atoms will conversely receive momentum along $z$-axis.

Clearly, the recoil momentum is transferred to the sample if the radioactive atoms are bound in it. Thus, for polarized nuclei, a recoil force emerges, acting on the sample as a whole. As was shown in Ref.~\cite{DeAngelis2012}, this force can be detected using an atomic force microscope. This is of interest because neutrino experiments are extremely complicated due to very small cross sections of weak processes. Therefore, it is useful to understand the possibilities of a new method to observe neutrinos, in particular, prospects to measure the neutrino mass as suggested in Ref.~\cite{Folan2014}. Note that a similar idea of using gravitational wave detector technology to measure and constrain particle interactions was discussed in Ref. \cite{Englert2018}.

Unfortunately, only one type of allowed transitions was considered in Ref.~\cite{DeAngelis2012}, a pure Gamow--Teller transition $J_i\to J_f=J_i-1$ ($\pi_i=\pi_f$), and the features of the weak interaction were accounted only qualitatively. Namely, for the transition specified, it was assumed that if the initial nuclei are completely polarized along $z$-axis, then all the neutrinos are emitted opposite to this axis (such a neutrino beam was named ''directed'').

In fact, the angular distribution of neutrinos (treated as massless particles) in EC for allowed nuclear transitions, as was first shown in Ref. \cite{Treiman1958}, is given by
\begin{equation}\label{2}
\frac{dW(\theta)}{d\Omega}=\frac{1}{4\pi}\left(1+BP\cos\theta\right),
\end{equation}
where $\theta$ is the angle between the neutrino momentum and the polarization axis ($z$-axis), $P$ is the polarization of the initial nuclei, $B$ is the asymmetry coefficient. Since $B=-1$ for the transition $J_i\to J_i-1$, then the neutrinos are emitted in all directions except $\theta=0$, if $P=1$; the angle $\theta=\pi$ corresponds to the maximum of the neutrino angular distribution. Thus, the direction opposite to the polarization axis is the most probable direction of neutrino emission, but not the only one. Therefore, the directionality of the neutrino beam was strongly overestimated in Ref.~\cite{DeAngelis2012}. This led, accordingly, to an overestimation of the average recoil momentum per neutrino emission and, consequently, of the recoil force acting on a sample.

The paper is organized as follows. In Section~\ref{s2} we briefly derive the angular distribution (\ref{2}) of neutrinos emitted in EC by polarized nuclei for the case of allowed nuclear transitions $J_i\to J_f=J_i, J_i\pm 1$ ($\pi_i=\pi_f$) taking into account the neutrino mass. In Section~\ref{s3} an expression for the recoil force acting on a radioactive sample is obtained. In particular, it is shown that in the case of $J_i\to J_i-1$ for completely polarized nuclei this force is three times smaller than that obtained from qualitative considerations in Ref. \cite{DeAngelis2012}. In Section~\ref{s4} electron-capturing isotopes suitable for neutrino recoil force measurement are presented. In Section~\ref{s5} we give numerical estimates for the recoil force. Due to some factors, the estimated values are smaller than those given in Ref. \cite{DeAngelis2012}. We show, nevertheless, that measurement of the recoil force seems possible by using the methods of magnetic resonance force microscopy. In Section~\ref{s6} we discuss characteristics of suitable isotopes and the corresponding recoil forces. Section~\ref{s7} is devoted to proposals for applying recoil force measurements in neutrino and weak interaction physics. In Conclusion (Section~\ref{s8}) we summarize the results and discuss possible ways to improve sensitivity of the proposed measurements.

\section{Neutrino emission asymmetry}\label{s2}

As mentioned above, the asymmetry of neutrino emission in EC was first obtained in Ref. \cite{Treiman1958}. Recently this result was reproduced in Ref. \cite{Vos2015a}, where some extra contributions were discussed resulted from a hypothetical Lorentz invariance violation. Below we present a short derivation of the differential rate for electron capture for the case of an allowed nuclear transition,  suggesting neutrino has a mass $m_{\nu}$, so that $E^2=p^2c^2+m_{\nu}^2c^4$, where $E=E_{\nu nx}$ and $p=p_{\nu nx}$ are the neutrino energy and momentum. 

Consider a polarized nucleus with spin $J_i\ge 1/2$ capture an electron from an arbitrary $x$-shell of the initial atom and undergo transition to the $n$-th state $|nJ_fM_f\rangle$ of the daughter nucleus with spin $J_f$ and its $z$-axis projection $M_f$. The initial state vector is a superposition
\begin{equation}\label{3}
|J_i\rangle=\sum_{M_i}a_{M_i}(J_i)|J_i M_i\rangle,\quad \sum_{M_i}|a_{M_i}(J_i)|^2=1,
\end{equation}
where summation is done over states with the $z$-axis projection $M_i$ of spin $J_i$. The neutrino emission rate in the direction $\mathbf{n}_{\nu}$ into a solid angle $d\Omega$ is determined by the Fermi golden rule:
\begin{multline}\label{4}
dw_{nEC}(\mathbf{n}_\nu) =
\sum_x dw_{nx}(\mathbf{n}_\nu)=\frac{2\pi}{\hbar} \\ \times 
\sum_{xM_f \sigma_e \sigma_\nu}
\left|\langle nJ_f M_f |\sum_j\hat h_{j}(\sigma_e, \sigma_\nu) |J_i\rangle \right|^2\frac{p E d\Omega}{(2\pi\hbar)^3c^2}\,,
\end{multline}
where 
\begin{multline}\label{5}
\hat h_{j}(\sigma_e, \sigma_{\nu}) = \frac{G_F V_{ud}}{\sqrt{2}}\,
e^{-i\efrac{\mathbf{p}\mathbf{r}_{j}}{\hbar}}\, \\ \times
\left[\,g_A\,\mathbf{j}(\sigma_e, \sigma_{\nu})\mbox{\boldmath $\sigma$}_j+
i g_V\, j_4(\sigma_e, \sigma_{\nu})\, \right] \hat\tau_{j-}
\end{multline}
is the weak interaction Hamiltonian, acting in the space of non-relativistic 2-component wave functions of the \linebreak$j$-th nucleon (summation over $j$ in Eq. (\ref{4}) is done over all nucleons); $\sigma_e$ and $\sigma_\nu$ are the spin projections on \linebreak $z$-axis for the captured electron and the emitted neutrino (we mainly use the notation from Ref. \cite[Ch. 10]{Eisenberg1976}; see also \cite{Barabanov1996,Barabanov2000}). Here $G_F$ is the Fermi constant; $V_{ud}$ is the element of quark mixing matrix; $g_V$ and $g_A$ are the vector and the axial nucleon form factors; $j_\lambda(\sigma_e, \sigma_{\nu})$ are the components of the lepton current four-vector ($\lambda = 1,2,3,4$); $\mathbf{r}_{j}$, $\mbox{\boldmath $\sigma$}_j$ and $\tau_{j-}$ are the position, spin operator and operator decreasing the isospin projection for the $j$-th nucleon. Since $pR/\hbar \ll 1$, where $R$ is the nuclear radius, in the following we take the exponential in (\ref{5}) equal to one.

The neutrino emission rate has to be averaged over the nuclear spin states. For $z$-axis chosen along the nuclear polarization vector, the spin density matrix averaged over the initial nuclei ensemble is diagonal,
\begin{equation}\label{6}
\langle a_{M_i}(J_i)a^*_{M_i^{\prime}}(J_i)\rangle=
\langle |a_{M_i}(J_i)|^2\rangle\, \delta_{M_i M_i^{\prime}}\,,
\end{equation}
and the nuclear polarization is given by
\begin{equation}\label{7}
P=\frac{\langle M_i\rangle}{J_i},\quad \langle M_i\rangle=\sum_{M_i}M_i \langle |a_{M_i}(J_i)|^2\rangle.
\end{equation}

Since our primary interest is in angular correlation, we use non-relativistic approximation for the captured electrons and treat the nucleus as point-like. Thus, the lepton current
\begin{equation}\label{8}
j_{\lambda}(\sigma_e, \sigma_\nu)=i u_{\nu}^{\dagger}(\sigma_{\nu}) \gamma_4 \gamma_{\lambda}
(1+\gamma_5) u_e(\sigma_e) \psi_x(0)
\end{equation}
is determined by the following bispinors:
\begin{align}\label{9}
u_{e}(\sigma_e) &= \left(\begin{array}{c}
\varphi_e (\sigma_e) \\ 0
\end{array}\right), \\
u_{\nu} (\sigma_{\nu})&=\sqrt{\frac{E+m_{\nu}c^2}{2E}}
\left(\begin{array}{c}
\varphi_{\nu} (\sigma_{\nu}) \\
\Frac{c\, \mbox{\boldmath $\sigma$} \mathbf{p}}{E+m_{\nu}c^2}\, \varphi_{\nu} (\sigma_{\nu}) \\
\end{array}\right),
\end{align}
where $\varphi_e,\,\varphi_{\nu} $ are two-component spinors, $\psi_x(0)$ is the radial wave function of the captured electron in the $x$-state at the origin (on the point-like nucleus). The phases of nuclear wave functions are chosen so that the matrix elements in Eq. (\ref{4}) are real \cite[\S 1B-2]{Bohr1998}. Using the Wigner--Eckart theorem, we express the matrix elements in terms of reduced matrix elements according to the definition given in \cite[Ch. 10]{Eisenberg1976},
\begin{multline}\label{10}
\langle nJ_fM_f|\sum_j \sigma_{jq} \hat\tau_{j-}|J_iM_i\rangle \\
= {\sqrt{\frac{2J_i+1}{2J_f+1}}}\,C^{J_fM_f}_{J_iM_i1q}\,M_{GT}(nJ_fJ_i),\\
\langle nJ_fM_f|\sum_j \hat\tau_{j-}|J_iM_i\rangle=
\delta_{J_fJ_i}\delta_{M_fM_i}\,M_F(nJ_iJ_i).
\end{multline}

\begin{figure}
\begin{center}
\includegraphics[scale=0.56]{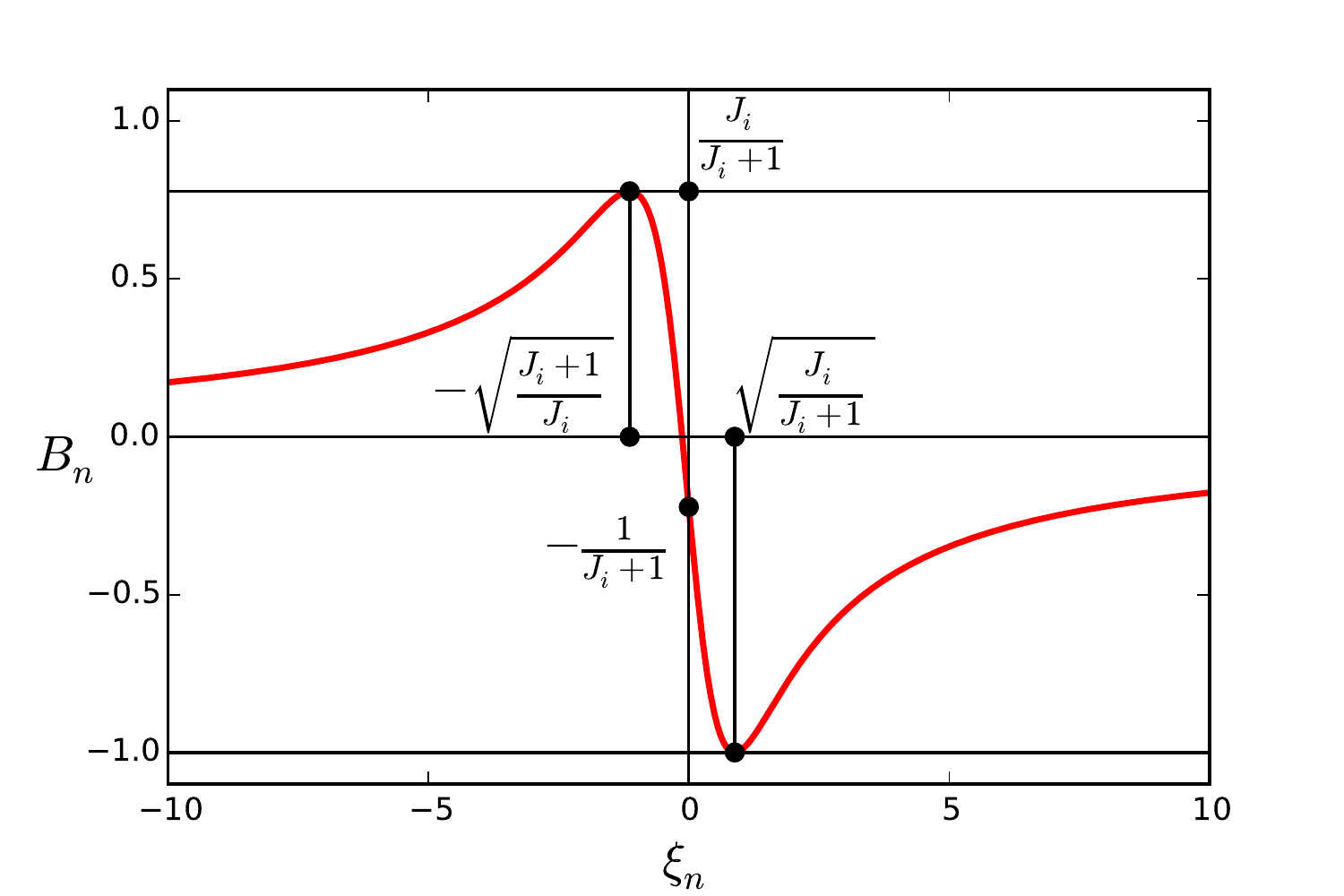}
\caption{The asymmetry coefficient $B_n$ as a function of the parameter $\xi_n$ (\ref{16}) for the case of mixed Fermi and Gamow--Teller transitions ($J_f=J_i$). Characteristic points are indicated. The curve is drawn for the value $J_i=7/2$.}\label{f1}
\end{center}
\end{figure}

To present the results, it is convenient to rewrite the neutrino emission rate (\ref{4}) and the corresponding angular distribution (\ref{2}) in the following form:
\begin{align}\label{12}
\frac{dw_{nEC}(\theta)}{d\Omega} &=\frac{w_{nEC}}{4\pi}\left(1+\eta_n B_nP\cos\theta\right),\\
\frac{dW(\theta)}{d\Omega} &=\frac{1}{w_{nEC}}\,\frac{dw_{nEC}(\theta)}{d\Omega},
\end{align}
where $w_{nEC}=\sum_x w_{nx}$ is the total rate of the transition $|J_i\rangle\to |nJ_f\rangle$ due to EC; the ratio $P_{nx}=w_{nx}/w_{nEC}$ determines the probability to capture an electron from the $x$-shell. The factor
\begin{equation}\label{12.1}
\eta_n=\frac{c\sum\limits_x p_{\nu nx}^2|\psi_x(0)|^2}{\sum\limits_x p_{\nu nx}E_{\nu nx}|\psi_x(0)|^2}
\end{equation}
is smaller than unity for $m_{\nu}\ne 0$ and becomes unity for massless neutrino. In the case of pure Gamow--Teller transitions $J_i\to J_f=J_i\pm 1$ (if $J_i=1/2$, then the only possible transition is $J_i=1/2\to J_f=J_i+1=3/2$), one obtains:
\begin{align}\label{13}
\begin{split}
w_{nEC} &=\phi_n\, g_A^2\, M^2_{GT}(nJ_fJ_i),\\
B_n &=\left\{\begin{array}{ll}
\Frac{J_i}{J_i+1}, & J_f=J_i+1,\\[\medskipamount]
-1, & J_f=J_i-1,
\end{array}\right.
\end{split}
\end{align}
where
\begin{equation}\label{14}
\phi_n=\frac{(G_F V_{ud})^2}{\pi\,\hbar^4c^2}\sum_x p_{\nu nx}E_{\nu nx}|\psi_x(0)|^2.
\end{equation}
If $J_f=J_i$, and a mixed Fermi and Gamow--Teller transition takes place, then
\begin{align}\label{15}
\begin{split}
w_{nEC} &=\phi_n 
\left[g_A^2\, M^2_{GT}(nJ_iJ_i)+g_V^2\,M^2_F(nJ_iJ_i)\right],\\
B_n &=-\frac{1+2\sqrt{J_i(J_i+1)}\,\xi_n}
{(J_i+1) (1+\xi_n^2)}\,,
\end{split}
\end{align}
where
\begin{equation}\label{16}
\xi_n=\frac{g_VM_F(nJ_iJ_i)}{g_AM_{GT}(nJ_iJ_i)}\,.
\end{equation}
The dependence of the asymmetry coefficient $B_n$ (\ref{15}) on the parameter $\xi_n$ (\ref{16}) with fixed $J_i$ is presented on Fig.~\ref{f1}. Note that the maximal (at $\xi_{n1}=-\sqrt{(J_i+1)/J_i}$\,) and the minimal (at $\xi_{n2}=\sqrt{J_i/(J_i+1)}$\,) values of $B_n$ coincide with the results for $J_f=J_i+1$ and $J_f=J_i-1$ given by Eq. (\ref{13}).

For massless neutrino our results coincide with those presented in Refs. \cite{Treiman1958,Vos2015a}. Clearly, the impact of neutrino mass is very small, so in what follows we neglect it except for the subsection specially devoted to the possibility to measure $m_{\nu}$.

\section{Neutrino recoil force}\label{s3}

Generally, an atom with a neutron-deficient nucleus is unstable with respect to EC and $\beta^+$-decay; let $I_{nEC}$ and $I_{n\beta+}$ be the corresponding branching ratios for transition to the $n$-th state of the final nucleus ($\sum_nI_{nEC}+\sum_nI_{n\beta+}=1$). The decay rate \linebreak $w_{nEC}=I_{nEC}\ln 2/T_{1/2}$ for the electron-capture transition is determined by the branching ratio $I_{nEC}$ and by the half-life $T_{1/2}$ of the radioactive atom. Note that a sample containing $N$ radioactive atoms has the activity $\alpha=N\ln 2/T_{1/2}$.

The $z$-component of recoil force $F_{nz}=\Delta P_{nz}/\Delta t$ is determined by the momentum
\begin{equation}\label{17}
\Delta P_{nz}= -N\Delta t\oint\frac{E_{\nu n}\cos\theta}{c}\, dw_{nEC}(\theta),
\end{equation}
transferred to the sample during the time $\Delta t$. The formula involves the neutrino energy averaged over atomic shells $x$,
\begin{equation}\label{17.2}
E_{\nu n}=\sum_x P_{nx} E_{\nu nx},
\end{equation}
so that $p_{nz}=E_{\nu n}\cos\theta/c$~ is the $z$-component of momentum for the neutrino emitted at the angle $\theta$. Substituting the differential rate~$dw_{nEC}(\theta)$ (\ref{12}) into Eq. (\ref{17}) and integrating over $d\Omega$, one obtains:
\begin{equation}\label{18}
F_{nz}= -\frac{N I_{nEC} \ln 2\, E_{\nu n} B_n P}{3\,c\, T_{1/2}}=
-\frac{\alpha I_{nEC} E_{\nu n} B_n P}{3\,c}\,.
\end{equation}
Thus, the recoil force, caused by the neutrino emission asymmetry, is determined by the product of the sample activity $\alpha$, the branching ratio $I_{nEC}$, the neutrino energy~$E_{\nu n}$, the asymmetry coefficient $B_n$, and the nuclear polarization $P$.

In a constant magnetic field $B$ at a temperature $T$, the polarization $P$ (\ref{7}) arises from the Boltzmann distribution of nuclear states (the $z$ axis is along $B$),
\begin{equation}\label{19b}
\langle|a_{M_i}(J_i)|^2\rangle\sim e^{-\efrac{E_{M_i}}{k_BT}}\,,
\end{equation}
where $E_{M_i}=-\mu B M_i/J_i$ is the state energy, $\mu$ is the nuclear magnetic moment, $k_B$ is the Boltzmann constant, and is given by (see, e.g., Ref. \cite{Kittel2005}):
\begin{equation}\label{20b}
P=\frac{2J_i+1}{2J_i}\,{\rm cth}\left(\frac{\beta (2J_i+1)}{2J_i}\right)-
\frac{1}{2J_i}\,{\rm cth}\left(\frac{\beta}{2J_i}\right),
\end{equation}
where $\beta=\mu B/(k_BT)$. Since nuclear magnetic moments are of the order of the nuclear magneton~$\mu_N$, the value of~$\beta$ is small even in a strong magnetic field $B$ at a relatively low temperature~$T$. Indeed, taking $B_0=1$~T and $T_0=1$~K one gets
\begin{equation}\label{21b}
\beta_0\equiv
\frac{\mu_N B_0}{k_BT_0}=3.658\cdot 10^{-4}.
\end{equation}
In the case of $\beta\ll 1$, the polarization (\ref{20b}) and the $z$-component of the sample magnetic moment $M_z=N\mu P$ take the form
\begin{equation}\label{22b}
P\simeq\frac{\beta (J_i+1)}{3J_i}\,,\quad
M_z\simeq\frac{NJ_i(J_i+1)\hbar^2\gamma^2 B}{3k_BT}\,,
\end{equation}
where $\gamma=\mu/(\hbar J_i)$ is the nuclear gyromagnetic ratio. As the temperature decreases, the magnetic moment (as well as the polarization) increases according to Curie's law: $M_z\sim 1/T$.

Assuming $\mu B\ll k_BT$ and using Eq. (\ref{22b}) for the polarization $P$, we rewrite the recoil force (\ref{18}) acting on a sample, which consists of one sort of radioactive atoms and has a mass $m$, as follows:
\begin{equation}\label{23b}
F_{nz}=-m\,\frac{B[T]}{T[K]}\,C_n f_n,
\end{equation}
where $B[T]$ is the magnetic field measured in Tesla (T), $T[K]$ is the temperature measured in Kelvin (K), the coefficient
\begin{equation}\label{24b}
C_n=B_n\,\frac{J_i+1}{J_i}
\end{equation}
is determined by the initial and final nuclear spins and by the transition type (see Eqs. (\ref{13}) and (\ref{15}) for $B_n$), while the force parameter 
\begin{equation}\label{25b}
f_n=\frac{\beta_0\, I_{nEC}\ln 2}{9\,T_{1/2}}\cdot
\frac{E_{\nu n}}{m_ac}\cdot\frac{\mu}{\mu_N},
\end{equation}
depends on the transition characteristics and the initial atom properties, in particular, on its mass~$m_a$.

The coefficient $B_n$ (as well as the coefficient $C_n$) can be either positive or negative; therefore, the contributions from different transitions, generally speaking, will partially cancel each other. Because of this, in the following we mainly discuss the simplest situation, when there is only one selected allowed transition with the branching ratio $I_{nEC}\ge 0.98$ (the contribution to the recoil force from neglected transitions will not exceed $\sim 2$\%, which is comparable with other uncertainties).

Pure Gamow--Teller transition is an especially simple case, for which the coefficient $B_n$ is known. If $J_i\to J_i-1$, then $B_n=-1$ (substitution into Eq. (\ref{18}) gives an expression which differs from the one obtained in Ref. \cite{DeAngelis2012} by the factor of $1/3$). For transitions $J_i\to J_i+1$, the coefficient~$B_n$ (\ref{13}) only slightly differs from unity, so the recoil force will be of the same order of magnitude as in the case of $J_i\to J_i-1$. As for the transitions of mixed type $J_i\to J_i$, the absolute value of the coefficient~$B_n$ can be comparable to unity at favorable values of parameter~$\xi_n$ (\ref{16}) (in principle, it is even possible to get $B_n=-1$ if $\xi_n=\xi_{n2}=\sqrt{J_i/(J_i+1)}$\,).

\section{Suitable isotopes}\label{s4}

A list of isotopes with non-zero spins decaying only ($I_{nEC}=1$) or mainly ($I_{nEC}\ge 0.98$) by EC via allowed transition to a single final nuclear state is not too wide. Recently we composed such a list for Gamow--Teller transitions in the context of the problem of modulated neutrino beams formation \cite{Barabanov2015} (see also \cite{Barabanov2017}). Here we present an extended list of suitable isotopes that decay via Gamow--Teller transition into the ground or excited states of daughter nuclei; see Tables~\ref{t1} and~\ref{t2}. As for transitions between the nuclei with the same (non-zero) spins and parities (i.e. for mixed Fermi and Gamow--Teller transitions), there are only two suitable isotopes (note that argon is a gas under standard conditions, but it solidifies at temperatures below 83.85~K); see the top part of Table~\ref{t3}. For both isotopes, transitions are going to the ground states of the final nuclei with the probability $I_{EC}=1$. Besides, we adduce three more isotopes with two transitions, to the ground and excited states of the final nucleus, with the total intensity $I_{EC}\ge 0.98$; see the bottom part of Table~\ref{t3}. One of the transitions is of pure Gamow-Teller type, while the other one is of mixed type.

\begin{table*}
\caption{\label{t1} List of Gamow--Teller transitions from the initial nucleus ${}^AX_i$ to the ground state ($n=0$) of the final nucleus ${}^AX_f$ due to EC. Here $T_{1/2}$ and $\mu$ are the half-life and the magnetic moment of the initial nucleus, $I_{n EC}$ is the branching ratio of the transition, $E_n^*$ is the excitation energy of the final nucleus, $Q_{n EC}$ is the energy release in EC, $E_{\nu n}$ is the neutrino energy, $f_n$ is the force parameter for the transition, $m_{\rm min}$, $N_{\rm min}$, $\alpha_{\rm min}$, $W_e^{\rm min}$, and $W_t^{\rm min}$ are the minimal values of the sample mass, number of decaying atoms, sample activity, heat load from secondary electrons, and total power of the electron-capture secondary products (see text for details), $P$ is the thermal polarization of the initial nuclei for $B=10$~T and $T=1$~K.}
\begin{center}
\setlength{\tabcolsep}{10pt}
{\renewcommand{\arraystretch}{0}%
\begin{ruledtabular}
\begin{tabular}{ccccccc}
\strut ${}^AX_i\,\to {}^AX_f {}^{\strut}_{\strut}$ &
$T_{1/2}$ &
$\mu/\mu_N$ &
$Q_{0 EC}\,\mbox{(keV)}$ &
$f_0\, \mbox{(N/g)}$ &
$N_{\rm min}$ &
$W_e^{\rm min}, W_t^{\rm min}\, \mbox{(nW)}$ \\
\strut $J_i^{\pi}\to J_f^{\pi} {}_{\strut}$ &
$I_{0 EC}\,(\%)$ &
$E^*_0\,\mbox{(keV)}$ &
$E_{\nu 0}\,\mbox{(keV)}$ &
$m_{\rm min}\, \mbox{(g)}$ &
$\alpha_{\rm min}\, \mbox{(MBq)}$ &
$P\,(\%) $ \\
\hline
\rule{0pt}{2pt} & & & & & & \\
%
%
\strut ${}^{163}\mbox{Er}\,\,\to {}^{163}\mbox{Ho} {}^{\strut}$ &
$75.0$~m &
$+0.557$ &
$1211$ &
$8.0\cdot 10^{-9}$ &
$4.6\cdot 10^9$ &
$0.73,\, 5.39$ \\
\strut $5/2^-\to 7/2^-$ &
$99.89$ &
$0$ &
$1164$ &
$1.3\cdot 10^{-12}$ &
$0.71$ &
$0.10$ \\
\rule{0pt}{2pt} & & & & & & \\
%
%
\strut ${}^{135}\mbox{La}\,\,\to {}^{135}\mbox{Ba} {}^{\strut}$ &
$19.5$~h &
$+3.70$ &
$1207$ &
$4.1\cdot 10^{-9}$ &
$1.6\cdot 10^{10}$ &
$0.15,\, 0.81$ \\
\strut $5/2^+\to 3/2^+$ &
$98.1$ &
$0$ &
$1175$ &
$3.6\cdot 10^{-12}$ &
$0.16$ &
$0.63$ \\
\rule{0pt}{2pt} & & & & & & \\
%
%
\strut ${}^{165}\mbox{Er}\,\,\to {}^{165}\mbox{Ho} {}^{\strut}$ &
$10.36$~h &
$+0.643$ &
$377$ &
$3.1\cdot 10^{-10}$ &
$1.2\cdot 10^{11}$ &
$1.85,\, 15.9$ \\
\strut $5/2^-\to 7/2^-$ &
$100$ &
$0$ &
$332$ &
$3.2\cdot 10^{-11}$ &
$2.16$ &
$0.11$ \\
\rule{0pt}{2pt} & & & & & & \\
%
%
\strut ${}^{131}\mbox{Cs}\,\,\to {}^{131}\mbox{Xe} {}^{\strut}$ &
$9.69$~d &
$+3.543$ &
$355$ &
$9.5\cdot 10^{-11}$ &
$3.4\cdot 10^{11}$ &
$0.23,\, 1.35$ \\
\strut $5/2^+\to 3/2^+$ &
$100$ &
$0$ &
$325$ &
$7.5\cdot 10^{-11}$ &
$0.29$ &
$0.60$ \\
\hline
\rule{0pt}{2pt} & & & & & & \\
%
%
\strut ${}^{71}\mbox{Ge}\,\,\to {}^{71}\mbox{Ga} {}^{\strut}$ &
$11.43$~d &
$+0.547$ &
$233$ &
$1.6\cdot 10^{-11}$ &
$5.4\cdot 10^{12}$ &
$2.87,\, 5.57$ \\
\strut $1/2^-\to 3/2^-$ &
$100$ &
$0$ &
$223$ &
$6.3\cdot 10^{-10}$ &
$3.77$ &
$0.20$ \\
\rule{0pt}{2pt} & & & & & & \\
%
%
\strut ${}^{55}\mbox{Fe}\,\,\to {}^{55}\mbox{Mn} {}^{\strut}$ &
$2.74$~y &
$+2.7$ &
$231$ &
$1.2\cdot 10^{-12}$ &
$9.5 \cdot 10^{13}$ &
$0.48,\, 0.71$ \\
\strut $3/2^-\to 5/2^-$ &
$100$ &
$0$ &
$225$ &
$8.6\cdot 10^{-9}$ &
$0.76$ &
$0.55$ \\
\rule{0pt}{2pt} & & & & & & \\
%
%
\strut ${}^{179}\mbox{Ta}\,\,\to {}^{179}\mbox{Hf} {}^{\strut}$ &
$1.82$~y &
$+2.289$ &
$106$ &
$1.4\cdot 10^{-13}$ &
$2.4 \cdot 10^{14}$ &
$2.08,\, 15.9$ \\
\strut $7/2^+\to 9/2^+$ &
$100$ &
$0$ &
$71$ &
$7.0\cdot 10^{-8}$ &
$2.84$ &
$0.36$ \\
\end{tabular}
\end{ruledtabular}
}
\end{center}
\end{table*}

In Tables~\ref{t1}-\ref{t3} we present the basic properties of the decaying nuclei (half-lives $T_{1/2}$, magnetic moments $\mu$) and of the selected nuclear transitions (branching ratio $I_{n EC}$, excitation energy of the final nucleus $E_n^*$, energy release $Q_{n EC}$, neutrino energy $E_{\nu n}$). These data are used to calculate the force parameter $f_n$ (its values are also shown in the Tables) and other relevant quantities. All numerical values are taken from the website \cite{IAEA}.

\begin{table*}
\caption{\label{t2} List of Gamow--Teller transitions from the initial nucleus ${}^AX_i$ to the excited $n$th state of the final nucleus ${}^AX_f^*$ due to EC. The quantities $T_{1/2}$, $\mu$, $I_{n EC}$, $E_n^*$, $Q_{n EC}$, $E_{\nu n}$, $f_n$, $m_{\rm min}$, $N_{\rm min}$, $\alpha_{\rm min}$, $W_e^{\rm min}$, $W_t^{\rm min}$, and $P$ are the same as in Table~\ref{t1}. Magnetic moments for ${}^{179}$W and ${}^{103}$Pd are unknown; they were taken to be $\mu_N$ as an estimate.}
\begin{center}
\setlength{\tabcolsep}{10pt}
{\renewcommand{\arraystretch}{0}%
\begin{ruledtabular}
\begin{tabular}{ccccccc}
\strut ${}^AX_i\,\to {}^AX^*_f {}^{\strut}_{\strut}$ &
$T_{1/2}$ &
$\mu/\mu_N$ &
$Q_{n EC}\, \mbox{(keV)}$ &
$f_n\, \mbox{(N/g)}$ &
$N_{\rm min}$ &
$W_e^{\rm min}, W_t^{\rm min}\, \mbox{(nW)}$ \\
\strut $J_i^{\pi}\to J_f^{\pi} {}_{\strut}$ &
$I_{n EC}\,(\%)$ &
$E_n^*\, \mbox{(keV)}$ &
$E_{\nu n}\, \mbox{(keV)}$ &
$m_{\rm min}\, \mbox{(g)}$ &
$\alpha_{\rm min}\, \mbox{(MBq)}$ &
$P\, (\%) $ \\
\hline
\rule{0pt}{2pt} & & & & & & \\
%
%
\strut ${}^{179}\mbox{W}\,\,\to {}^{179}\mbox{Ta}^* {}^{\strut}$ &
$37.05$~m &
$(1)$  &
$1032$ &
$2.2\cdot 10^{-8}$ &
$1.5\cdot 10^{9}$ &
$1.17,\, 6.60$ \\
\strut $7/2^-\to 9/2^-$ &
$99.2$ &
$30.7$ &
$975$ &
$4.5\cdot 10^{-13}$ &
$0.48$ &
$0.16$ \\
\rule{0pt}{2pt} & & & & & & \\
%
%
\strut ${}^{107}\mbox{Cd}\,\,\to {}^{107}\mbox{Ag}^* {}^{\strut}$ &
$6.50$~h &
$-0.615$ &
$1323$ &
$2.9\cdot 10^{-9}$ &
$2.0\cdot 10^{10}$ &
$7.91,\, 10.7$ \\
\strut $5/2^+\to 7/2^+$ &
$99.7$ &
$93.1$ &
$1301$ &
$3.5\cdot 10^{-12}$ &
$0.58$ &
$0.10$ \\
\rule{0pt}{2pt} & & & & & & \\
%
%
\strut ${}^{119}\mbox{Sb}\,\,\to {}^{119}\mbox{Sn}^* {}^{\strut}$ &
$38.2$~h &
$+3.450$ &
$567$ &
$1.0\cdot 10^{-9}$ &
$3.5\cdot 10^{10}$ &
$0.61,\, 1.39$ \\
\strut $5/2^+\to 3/2^+$ &
$100$ &
$23.9$ &
$542$ &
$6.9\cdot 10^{-12}$ &
$0.18$ &
$0.59$ \\
\rule{0pt}{2pt} & & & & & & \\
%
%
\strut ${}^{111}\mbox{In}\,\,\to {}^{111}\mbox{Cd}^* {}^{\strut}$ &
$2.805$~d &
$+5.503$ &
$443$ &
$7.8\cdot 10^{-10}$ &
$5.7\cdot 10^{10}$ &
$0.84,\, 11.5$ \\
\strut $9/2^+\to 7/2^+$ &
$100$ &
$416.6$ &
$420$ &
$1.1\cdot 10^{-11}$ &
$0.16$ &
$0.82$ \\
\hline
\rule{0pt}{2pt} & & & & & & \\
%
%
\strut ${}^{103}\mbox{Pd}\,\,\to {}^{103}\mbox{Rh}^* {}^{\strut}$ &
$16.99$~d &
$(1)$ &
$535$ &
$3.1\cdot 10^{-11}$ &
$1.9\cdot 10^{12}$ &
$5.77,\, 8.62$ \\
\strut $5/2^+\to 7/2^+$ &
$99.9$ &
$39.7$ &
$514$ &
$3.2\cdot 10^{-10}$ &
$0.90$ &
$0.17$ \\
\rule{0pt}{2pt} & & & & & & \\
%
%
\strut ${}^{57}\mbox{Co}\,\,\to {}^{57}\mbox{Fe}^* {}^{\strut}$ &
$271.7$~d &
$+4.720$ &
$700$ &
$2.2\cdot 10^{-11}$ &
$3.7\cdot 10^{12}$ &
$0.31,\, 2.51$ \\
\strut $7/2^-\to 5/2^-$ &
$99.8$ &
$136.5$ &
$692$ &
$3.5\cdot 10^{-10}$ &
$0.11$ &
$0.74$ \\
\rule{0pt}{2pt} & & & & & & \\
%
%
\strut ${}^{54}\mbox{Mn}\,\,\to {}^{54}\mbox{Cr}^* {}^{\strut}$ &
$312.2$~d &
$+3.282$ &
$542$ &
$1.1\cdot 10^{-11}$ &
$7.6\cdot 10^{12}$ &
$0.13,\, 26.4$ \\
\strut $3^+\to 2^+$ &
$100$ &
$834.8$ &
$537$ &
$6.8\cdot 10^{-10}$ &
$0.20$ &
$0.53$ \\
\rule{0pt}{2pt} & & & & & & \\
%
%
\strut ${}^{73}\mbox{As}\,\,\to {}^{73}\mbox{Ge}^* {}^{\strut}$ &
$80.30$~d &
$+1.63$ &
$278$ &
$7.8\cdot 10^{-12}$ &
$6.3\cdot 10^{12}$ &
$5.82,\, 7.75$ \\
\strut $3/2^-\to 1/2^-$ &
$100$ &
$66.7$ &
$268$ &
$7.7\cdot 10^{-10}$ &
$0.63$ &
$0.33$ \\
\rule{0pt}{2pt} & & & & & & \\
%
%
\strut ${}^{125}\mbox{I}\,\,\to {}^{125}\mbox{Te}^* {}^{\strut}$ &
$59.41$~d &
$2.821$ &
$150$ &
$5.0\cdot 10^{-12}$ &
$7.0\cdot 10^{12}$ &
$2.49,\, 9.3$ \\
\strut $5/2^+\to 3/2^+$ &
$100$ &
$35.5$ &
$124$ &
$1.4\cdot 10^{-9}$ &
$0.94$ &
$0.48$ \\
\rule{0pt}{2pt} & & & & & & \\
%
%
\strut ${}^{139}\mbox{Ce}\,\,\to {}^{139}\mbox{La}^* {}^{\strut}$ &
$137.63$~d &
$1.06$ &
$113$ &
$4.9\cdot 10^{-13}$ &
$8.9\cdot 10^{13}$ &
$26.7,\, 162.7$ \\
\strut $3/2^+\to 5/2^+$ &
$100$ &
$165.9$ &
$84$ &
$2.1\cdot 10^{-8}$ &
$5.21$ &
$0.22$ \\
\rule{0pt}{2pt} & & & & & & \\
%
%
\strut ${}^{109}\mbox{Cd}\,\,\to {}^{109}\mbox{Ag}^* {}^{\strut}$ &
$461.9$~d &
$-0.828$ &
$127$ &
$1.8\cdot 10^{-13}$ &
$3.0\cdot 10^{14}$ &
$68.4,\, 92.1$ \\
\strut $5/2^+\to 7/2^+$ &
$100$ &
$88.0$ &
$106$ &
$5.5\cdot 10^{-8}$ &
$5.26$ &
$0.14$ \\
\end{tabular}
\end{ruledtabular}
}
\end{center}
\end{table*}

Note, that the values of the neutrino energy $E_{\nu n}$ (\ref{17.2}) are provided on the website \cite{IAEA} for some isotopes, but not for all. However, the values of $E_{\nu n}$ can be easily found. Indeed, the energies $E_{\nu nx}$ (\ref{1}) follow from the data~\cite{Bearden1967} on excitation energies $E_x$ of the final-state atoms with holes in $x$ shells, while the probabilities $P_{nx}$ are given on \cite{IAEA} for $x=K$ and $L$ for all transitions. Thus, we calculated the energies $E_{\nu n}$ for all isotopes of interest using Eq. (\ref{17.2}) and assuming $P_{nM}=1-P_{nK}-P_{nL}$. The reliability of the calculated values is confirmed by their agreement with the ones presented on the website~\cite{IAEA}.

\begin{table*}
\caption{\label{t3} List of mixed Fermi and Gamow--Teller transitions from the initial nucleus ${}^AX_i$ to the $n$th state of the final nucleus ${}^AX_f$ due to EC (for isotopes in the bottom of the Table, pure Gamow--Teller transitions take place as well; see text for details). The quantities $T_{1/2}$, $\mu$, $I_{n EC}$, $E_n^*$, $Q_{n EC}$, $E_{\nu n}$, $f_n$ are the same as in Table~\ref{t1}; $m$, $N$, $\alpha$, $W_e$, and $W_t$ are the sample mass, number of decaying atoms, sample activity, heat load from secondary electrons, and total power of the electron-capture secondary products (see text for details), $F_n$ is the recoil force for pure Gamow--Teller transition or the maximal recoil force for mixed Fermi and Gamow--Teller transition.}
\begin{center}
\setlength{\tabcolsep}{10pt}
{\renewcommand{\arraystretch}{0}%
\begin{ruledtabular}
\begin{tabular}{ccccccc}
\strut ${}^AX_i\,\to {}^AX_f {}^{\strut}_{\strut}$ &
$T_{1/2}$ &
$\mu/\mu_N$ &
$m\, \mbox{(g)}$ &
$N$ &
$\alpha\, \mbox{(MBq)}$ &
$W_e, W_t \, \mbox{(nW)}$ \\
\strut $J_i^{\pi}\to J_f^{\pi} {}_{\strut}$ &
$I_{n EC}\,(\%)$ &
$E_n^*\, \mbox{(keV)}$ &
$Q_{n EC}\, \mbox{(keV)}$ &
$E_{\nu n}\, \mbox{(keV)}$ &
$f_n\, \mbox{(N/g)}$ &
$F_n\, \mbox{(N)}$ \\
\hline
\rule{0pt}{2pt} & & & & & & \\
%
%
\strut ${}^{37}\mbox{Ar}\,\,\to {}^{37}\mbox{Cl} {}^{\strut}$ &
$35.01$~d &
$+1.145$ &
$1.0\cdot 10^{-10}$ &
$1.6\cdot 10^{12}$ &
$0.37$ &
$0.12,\, 0.15$ \\
\strut $3/2^+\to 3/2^+$ &
$100$ &
$0$ &
$814$ &
$811$ &
$7.5\cdot 10^{-11}$ &
$1.3\cdot 10^{-19}$ \\
\rule{0pt}{2pt} & & & & & & \\
%
%
\strut ${}^{49}\mbox{V}\,\,\to {}^{49}\mbox{Ti} {}^{\strut}$ &
$330$~d &
$4.47$ &
$1.0\cdot 10^{-9}$ &
$1.2\cdot 10^{13}$ &
$0.30$ &
$0.16,\, 0.21$ \\
\strut $7/2^-\to 7/2^-$ &
$100$ &
$0$ &
$602$ &
$597$ &
$1.7\cdot 10^{-11}$ &
$2.2\cdot 10^{-19}$ \\
\hline
\rule{0pt}{2pt} & & & & & & \\
%
%
\strut ${}^{7}\mbox{Be}\,\,\to {}^{7}\mbox{Li}^*, {}^{7}\mbox{Li} {}^{\strut}$ &
$53.22$~d &
$-1.399$ &
$1.0\cdot 10^{-10}$ &
$8.6\cdot 10^{12}$ &
$1.29$ &
$0.00,\, 10.4$ \\
$\begin{array}{c} \strut 3/2^-\to 1/2^- \\ \strut 3/2^-\to 3/2^- \end{array}$ &
$\begin{array}{c} \strut 10.44 \\ \strut 89.56 \end{array}$ &
$\begin{array}{c} \strut 477.6 \\ \strut 0 \end{array}$ &
$\begin{array}{c} \strut 384 \\ \strut 862 \end{array}$ &
$\begin{array}{c} \strut 384 \\ \strut 862 \end{array}$ &
$\begin{array}{c} \strut 1.6\cdot 10^{-11} \\ \strut 3.0\cdot 10^{-10} \end{array}$ &
$\begin{array}{c} \strut 0.3\cdot 10^{-19} \\ \strut 5.1\cdot 10^{-19} \end{array}$ \\
\rule{0pt}{2pt} & & & & & & \\
%
%
\strut ${}^{51}\mbox{Cr}\,\,\to {}^{51}\mbox{V}^*, {}^{51}\mbox{V} {}^{\strut}$ &
$27.70$~d &
$-0.93$ &
$1.0\cdot 10^{-9}$ &
$1.2\cdot 10^{13}$ &
$3.42$ &
$2.00,\, 20.1$ \\
$\begin{array}{c} \strut 7/2^-\to 5/2^- \\ \strut 7/2^-\to 7/2^- \end{array}$ &
$\begin{array}{c} \strut 9.93 \\ \strut 90.07 \end{array}$ &
$\begin{array}{c} \strut 320.1 \\ \strut 0 \end{array}$ &
$\begin{array}{c} \strut 432 \\ \strut 752 \end{array}$ &
$\begin{array}{c} \strut 427 \\ \strut 748 \end{array}$ &
$\begin{array}{c} \strut 3.0\cdot 10^{-12} \\ \strut 4.7\cdot 10^{-11} \end{array}$ &
$\begin{array}{c} \strut 0.4\cdot 10^{-19} \\ \strut 6.0\cdot 10^{-19} \end{array}$ \\
\rule{0pt}{2pt} & & & & & & \\
%
%
\strut ${}^{65}\mbox{Zn}\,\,\to {}^{65}\mbox{Cu}^*, {}^{65}\mbox{Cu} {}^{\strut}$ &
$243.9$~d &
$+0.769$ &
$1.0\cdot 10^{-7}$ &
$9.3\cdot 10^{14}$ &
$30.50$ &
$22.6,\, 2766.4$ \\
$\begin{array}{c} \strut 5/2^-\to 5/2^- \\ \strut 5/2^-\to 3/2^- \end{array}$ &
$\begin{array}{c} \strut 50.04 \\ \strut 48.54 \end{array}$ &
$\begin{array}{c} \strut 1115.6 \\ \strut 0 \end{array}$ &
$\begin{array}{c} \strut 236 \\ \strut 1352 \end{array}$ &
$\begin{array}{c} \strut 228 \\ \strut 1344 \end{array}$ &
$\begin{array}{c} \strut 5.8\cdot 10^{-13} \\ \strut 3.3\cdot 10^{-12} \end{array}$ &
$\begin{array}{c} \strut 8.1\cdot 10^{-19} \\ \strut 46.5\cdot 10^{-19} \end{array}$ \\
\end{tabular}
\end{ruledtabular}
}
\end{center}
\end{table*}

We also performed similar computations for the energies of electron-capture secondary products and the corresponding heat loads. If the transition occurs to the ground state of the nucleus, the total released energy is distributed between the neutrino and the final-state atom excitation: $\Delta (mc^2)=E_{\nu 0x}+E_x$ (see Eq.~(\ref{1}); as mentioned previously, we neglect the recoil energy). The energy $E_x$ is, in turn, distributed between the secondary products of the decay, ${\rm x}$-rays and Auger electrons (AE), emitted by the excited atom. For the transition to an excited state of the final nucleus, we have $\Delta (mc^2)=E_{\nu nx}+E_n^*+E_x$. The nuclear excitation energy $E_n^*$ is also distributed between the secondary products: $\gamma$-rays, conversion electrons (CE) and additional ${\rm x}$-rays and AE resulted from recombination of holes in the $K$-, $L$-, $M$-,\,\ldots\, shells of the final atom, caused by CE emission. The total energy of the secondary products per decay is given by
\begin{equation}\label{25.3b}
E_t=E^*_n+\sum_xP_{nx}E_x,
\end{equation}
and the corresponding total power caused by EC and released in a sample is
\begin{equation}\label{25.2b}
W_t=\alpha I_{n EC} E_t.
\end{equation}

However, it is crucial to know the energy $E_{\rm AE,CE}$ released in electron emission. Indeed, as it was reasonably noted in Ref. \cite{DeAngelis2012}, the experimental setup can be designed so that the radiation will not be absorbed in the radioactive sample or in the cooled area. Thus, the heat load will be mainly determined by the secondary AE and CE. Using integral and differential data on energies and intensities for secondary processes presented on the website~\cite{IAEA}, we found the values of $E_{\rm AE,CE}$ and the corresponding heat loads
\begin{equation}\label{25.2c}
W_e=\alpha I_{n EC} E_{\rm AE,CE}
\end{equation}
for all the isotopes and transitions of interest.

\section{Neutrino recoil force: measurement prospects}\label{s5}

The key element of an atomic force microscope is the cantilever, a micromechanical beam of length $l$, width $w$ and thickness $t$, made of a material with Young's modulus $E$ clamped at one end and with a tip at the other one (see, e.g., \cite{Hammel2007}). The force $F$ acting on the tip and its displacement $z$ are related by Hooke's law $z=F/k$, where the spring constant is given by
\begin{equation} \label{19}
k\simeq \frac{E w t^3}{4 l^3}\,.
\end{equation}
Atomic force microscope is usually operated in either contact or noncontact mode. In contact mode, the cantilever is in hard contact with the surface and moves over it; measuring the tip displacement gives the force acting on it. In noncontact mode, the tip on the free end oscillates with the fundamental frequency of the cantilever~$\omega_c$, while placed at a distance from the surface; this allows to determine, e.g., a force gradient by measuring the frequency shift.

In Ref. \cite{Ohnesorge1993}, an atomic force microscope with the cantilever spring constant 0.2~N/m was operated in contact mode, and the accuracy of force measurement was $10^{-12}$~N. With this result, the authors of \cite{DeAngelis2012} estimated the mass of a $^{119}$Sb sample, required to obtain the neutrino recoil force $F=10^{-12}$~N at 100\% nuclear polarization. Reproducing this estimate by Eq. (\ref{18}) with $B_n=-1$, $E_{\nu n}=542$~keV and $I_{n EC}=1$ (see Table~\ref{t2}), we obtain the activity $\alpha=10.4$~GBq, the number of atoms $N=2.1\cdot 10^{15}$ and the sample mass $m=4.1\cdot 10^{-7}$~g (our results for $\alpha$, $N$ and $m$ are three times greater than those from \cite{DeAngelis2012}, because, as previously mentioned, our formula for the recoil force contains an additional factor of 1/3). The mass of the sample is comparable to that of the silicon cantilever tip $m_t=1.4\cdot 10^{-7}$~g, calculated by the authors of \cite{DeAngelis2012} from the data given in Ref.~\cite{Ohnesorge1993}. Therefore, if the polarization of $^{119}$Sb nuclei can be raised to unity by the use of extremely strong intra-atomic magnetic field at very low temperature (as it was assumed in Ref. \cite{DeAngelis2012}), so that $\mu B\sim k_BT$ (we discuss this option below), the situation for the antimony isotope seems quite optimistic. If, however, one considers the case of $\mu B\ll k_BT$ with a relatively low nuclear polarization, the recoil force of $10^{-12}$~N appears inaccessible for $^{119}$Sb as well as for other isotopes.

Fortunately, the use of an oscillating cantilever in noncontact mode allows to measure forces much smaller than $10^{-12}$~N. In particular, this is the case for magnetic resonance force microscopy (MRFM) \cite{Hammel2007,Sidles1995,Suter2004,Greenberg2012}. In one of the versions of this method a sample with a magnetic moment $M_z$ is attached to a cantilever; the force acting on the sample results from a gradient $\nabla B(z)$ of inhomogeneous magnetic field. Note, that magnetic moment of the sample is due either to unpaired electrons or to nuclei with non-zero spins (and magnetic moments). Thus, the methods of electron paramagnetic resonance (EPR) or nuclear magnetic resonance (NMR) can be applied. Namely, affecting the sample by a specifically modulated oscillating magnetic field (for NMR, with a frequency $\omega_{{\rm rf}}$ close to $\omega_{{\rm NMR}}=\gamma B$), one induces oscillations of magnetic moment $M_z$ with the modulation frequency $\omega$. In the case of NMR, one uses the method of cyclic adiabatic inversion (see details in Ref. \cite{Suter2004}). Then the force
\begin{equation}\label{19.2}
F_z=M_z\nabla B(z)
\end{equation}
oscillates with the same frequency $\omega$. When $\omega=\omega_c$, we get a resonance at which the amplitude of the cantilever displacement caused by the force of amplitude $F_0$ reaches its maximal value $x_0=QF_0/k$, where $Q$ is the quality factor. Hence, with a fixed accuracy of displacement measurement, the sensitivity to the force increases by a factor of $Q$. This is one of the methods of magnetic resonance registration, used in MRFM. The advantage of the method is its high sensitivity: magnetic resonance is detected in very small samples, for which the standard registration methods are inapplicable.

For our purposes, the following is important. Let us assume that a sample consisting of $N$ electron-capturing atoms and attached to a cantilever is put in a constant and homogeneous magnetic field $B$. The nuclear polarization $P$ in equilibrium is given by Eq. (\ref{20b}) or, in the case of $\mu B\ll k_BT$, by Eq. (\ref{22b}). Using cyclic adibatic inversion, one can initiate oscillations of the nuclear magnetic moment $M_z=N\mu P$ of the sample at the resonant frequency of the cantilever $\omega_c$. But these oscillations are, in fact, the oscillations of polarization $P$. Therefore, the neutrino recoil force (\ref{18}), proportional to $P$, will also oscillate. In this case, Eqs. (\ref{18}) and (\ref{23b}) determine the amplitude $F_n=|F_{nz}|$ of this force.

Thus, the neutrino recoil force can be measured in the same manner as the force acting on a magnetized sample in MRFM. Of course, the homogeneity of the magnetic field has to be sufficiently high to ensure that the magnetic force (\ref{19.2}) is much smaller than the recoil force.

The limitations of the method described above are related primarily to the thermal fluctuations \cite{Suter2004}. At a given temperature $T$, the minimally measurable force is
\cite[Eq. (4.10a)]{Sidles1995} (see also \cite{Suter2004,Stowe1997}):
\begin{equation}\label{26b}
F_{\rm min}=\sqrt{\frac{4kk_BT\Delta\nu}{Q\omega_c}}\,,
\end{equation}
where $\Delta\nu$ is the measurement bandwidth. For estimates, let us assume it equal to the half width at half maximum of the resonance, $\Delta\nu=\omega_c/(2Q)$ (this is equivalent to a requirement $\Delta\nu=1/\tau$, where $\tau=2Q/\omega_c$~is the oscillator damping time); this leads to
\begin{equation}\label{27b}
F_{\rm min}=\frac{\sqrt{2kk_BT}}{Q}\,.
\end{equation}
Evidently, the sensitivity to the force can be improved by increasing the quality factor~$Q$, lowering the temperature $T$ and reducing the spring constant $k$ (\ref{19}).

\begin{figure}
\begin{center}
\includegraphics[scale=0.6]{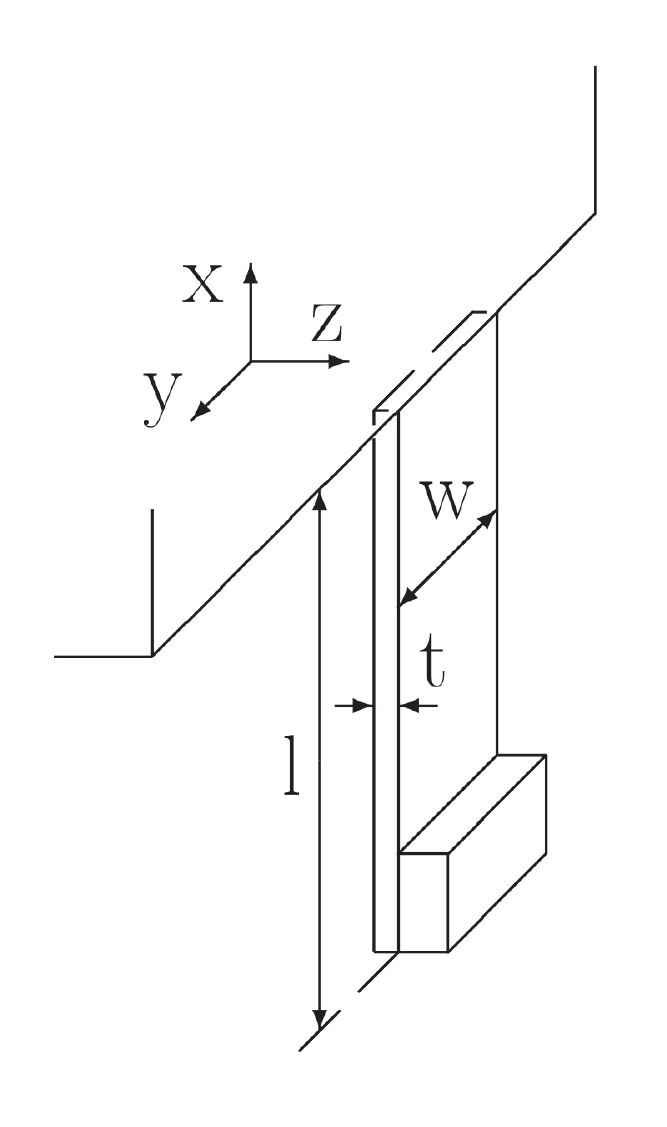}
\caption{Scheme of a micromechanical resonator with a mass load on the tip; $l$, $w$ and $t$ are the length, the width and the thickness of the resonator.}\label{f2}
\end{center}
\end{figure}

In Ref.~\cite{Stowe1997}, a technology was presented to produce thin (up to $t=50$~nm) and long (up to $l=400$~$\mu$m) cantilevers, made of single-crystal silicon with a spring constant up to $10^{-5}$~N/m and a quality factor of $10^3$--$10^4$; further development allowed to achieve $Q\sim 10^5$ \cite{Hammel2007,Suter2004}. Since
\begin{equation}\label{27.2b}
F_{\rm min}\simeq 10^{-19}~\mbox{N}
\end{equation}
at $k=10^{-5}$~N/m, $T=1$~K and $Q=10^5$, this technology opened a possibility to measure attonewton and sub-attonewton forces. In addition, a mass load to the free end of a cantilever to suppress oscillation modes of high orders was proposed in Ref.~\cite{Mozyrsky2003}. Note that the cantilever is positioned vertically; its upper end is clamped (see Fig.~\ref{f2}). Ultra-thin cantilevers of this type with the mass load slightly exceeding the mass of the cantilever were successfully used, e.g., in studies~\cite{Mamin2005,Mamin2007,Xue2011}. The mass load determines, in fact, the effective mass of the oscillator $m_{{\rm eff}}$, which, along with the spring constant $k$, gives the fundamental frequency of the cantilever oscillations:
\begin{equation}\label{28b}
\omega_c=\sqrt{\frac{k}{m_{\rm eff}}}\,.
\end{equation}

To obtain the estimate (\ref{27.2b}) we used the value $T=1$~K taking into account the following. In order to enhance the quality factor, modern studies in the field of MRFM are performed at low temperatures (and at high vacuum). However, in the process of sample remagnetization, during cyclic adiabatic inversion, heat is produced. The system also receives energy from the optical laser interferometer, which is used to detect the cantilever deflection. As the result of efforts to minimize the heat load, the temperature was lowered to $T\simeq 1$~K \cite{Xue2011} (in particular, the laser light power in the interferometer was reduced to 100~nW). This is why below we keep the ''conservative'' estimate $T=1$~K for the temperature.

Let a radioactive sample of mass $m$ be a mass load. What is the upper limit on $m$? Following Ref.~\cite{Stowe1997}, we take a cantilever with a maximal length $l=400$~$\mu$m and a maximal reasonable width $w=l/10$ made of single-crystal silicon (with Young's modulus $E=1.31$~GPa and the density $\rho=2.33$~g/cm$^3$). According to Eq.~(\ref{19}), the required value of $k$ corresponds to the thickness $t\simeq 80$~nm and, therefore, the cantilever mass $m_c\simeq 0.3\cdot 10^{-8}$~g. Taking the density $\rho=5$~g/cm$^3$ and the volume of $40\times 40\times 10$~($\mu$m)$^3$ for the sample, we obtain $m_{\rm max}\simeq 10^{-7}$~g for the maximal mass. The oscillation frequency (\ref{28b}) of such a loaded cantilever is $\nu_c=\omega_c/(2\pi)\simeq 50$~Hz (note that usually kHz-range is used, however, e.g., the MRFM based study~\cite{Thurber2003} was performed with $\nu_c=490$~Hz).
 
Obviously, for a small sample with a mass, say, $m\simeq m_0=10^{-10}$~g, a better option is a cantilever, made of the same material and with the same spring constant as discussed above, but with smaller dimensions $l=100$~$\mu$m, $w=2$~$\mu$m, $t=50$~nm and the mass $m_c\simeq 2.3\cdot 10^{-11}$~g (similar cantilevers with a mass load $\sim 10^{-10}$~g were used in Refs. \cite{Mamin2005,Mamin2007,Xue2011}). The oscillation frequency of such a ''small'' cantilever is $\nu_{c0}\simeq 1.5$~kHz. For definiteness, let us assume that a radioactive sample of mass $m<m_0$ is placed on the ''small'' cantilever, which is additionally loaded up to the mass $m_0$ with any non-radioactive material (so its frequency is still $\nu_{c0}$).

Applying the obtained limits on $F_n$ and $m$ to Eq. (\ref{23b}), we get
\begin{equation}\label{29b}
\frac{F_n}{m}=
\frac{B[T]}{T[K]}\,|C_n|\,f_n\ge
\frac{F_{\rm min}}{m}\ge
\frac{F_{\rm min}}{m_{\rm max}}\simeq 10^{-12}\,\,
\frac{\rm N}{\rm g}\,.
\end{equation}
Thus, the method described above allows to measure the neutrino recoil force for a sample of electron-capturing isotopes provided the force parameter $f_n$ satisfies the condition (\ref{29b}) for reasonable values of magnetic field $B$ and temperature $T$.

\section{Neutrino recoil force for the selected isotopes}\label{s6}

Taking into account the capability of modern superconducting magnets, we suppose $B=10$~T. The absolute value of the coefficient~$C_n$ (\ref{24b}) is close to unity for pure Gamow--Teller transitions, as well as for mixed transitions with a favorable value of the parameter~$\xi_n$. Thus, setting $|C_n|\simeq 1$ and $T=1$~K, we obtain from (\ref{29b}):
\begin{equation}\label{30b}
f_n\ge 10^{-13}\,\,
\frac{\rm N}{\rm g}\,.
\end{equation}
We used this condition to select the isotopes for Tables~\ref{t1}-\ref{t3}. For many transitions in the Tables the condition is fulfilled with a margin.

Let us find the minimal sample mass providing a detectable neutrino recoil force for each of the selected Gamow--Teller transitions (see Tables~\ref{t1} and~\ref{t2}). To do this we substitute $B=10$~T and $T=1$~K into Eq. (\ref{23b}) and rewrite it in the form:
\begin{equation}\label{31b}
F_n=10m|C_n|f_n\ge F_{\rm min}
\,\Rightarrow\,
m\ge \frac{F_{\rm min}}{10|C_n|f_n}\equiv m_1.
\end{equation}
There is, however, an additional lower limit on the sample mass: during one period of cantilever oscillations, the average number of emitted neutrinos should be sufficiently large. Taking this number equal 100 (as an estimate), we get
\begin{equation}\label{32b}
\frac{\alpha}{\nu_c}=
\frac{m\,\ln 2}{m_a\,T_{1/2}\,\nu_c}\ge 100\,.
\end{equation}
The frequency $\nu_c=\omega_c/(2\pi)$ is determined by Eq. (\ref{28b}), where $m_{\rm eff}=m$, if $m>m_0=10^{-10}$~g, and $m_0$, if $m<m_0$. In practice, it turned out that $m_{\rm min}=m_1$ for all isotopes, except for the sample of $^{135}$La (its minimal mass was found from Eq. (\ref{32b})). Clearly, an increase in the mass $m$ (to increase the recoil force) above the minimal value $m_{\rm min}$ certainly results in the raise of $\alpha/\nu_c$.

In Tables~\ref{t1}-\ref{t2} for each isotope (and each selected transition) we present the minimal sample mass $m_{\rm min}$ and the corresponding number of radioactive atoms $N_{\rm min}$, sample activity $\alpha_{\rm min}$, heat powers caused by the electron-capture secondary products $W_e^{\rm min}$ and $W_t^{\rm min}$, as well as the polarization $P$ of the initial nuclei. The isotopes in the Tables are arranged in the descending order of the force parameter $f_n$ that corresponds, as one can see, to the ascending order for the minimal mass $m_{\rm min}$. The top and the bottom parts of the Tables are separated by a line: in the top part $m_{\rm min}<m_0$, while in the bottom part $m_{\rm min}>m_0$.

Notice that the force parameter $f_n$ (\ref{25b}) depends on four quantities: the half-life $T_{1/2}$ and the magnetic moment $\mu$ of the initial nucleus, the mass of the initial atom $m_a$ and the neutrino energy (averaged over $x$) $E_{\nu n}$. All these quantities significantly vary from nucleus to nucleus and from transition to transition. Nevertheless, in the case of one predominant transition there is a correlation between the half-life $T_{1/2}$ and the energy $E_{\nu n}$ (which is slightly less than $Q_{n EC}$), because the larger is $E_{\nu n}$, the less is $T_{1/2}$. Hence, the ratio $E_{\nu n}/T_{1/2}$ in Eq. (\ref{25b}) is decisive. So it is not surprising that the isotopes (and transitions) in Tables~\ref{t1} and~\ref{t2} are arranged in such a way that with the decrease of $f_n$ the value of $T_{1/2}$ tends to increase, while the energy $E_{\nu n}$ tends to decrease. Variations in the magnetic moment $\mu$ and the mass number (along with the mass $m_a$) bring in some irregularities in these tendencies. Similarly, with the increase of $m_{\rm min}$ the total number of atoms $N_{\rm min}$ increases almost monotonically.

For nuclei with mixed Fermi and Gamow--Teller transitions (see Table~\ref{t3}) we use another approach. Here it is possible to determine only the maximal force, taking $|C_n|=|C_n^{\rm max}|=(J_i+1)/J_i$ in Eq. (\ref{23b}). For each isotope in Table~\ref{t3} the sample mass $m$ was chosen from the rounded-off values $10^{-7}$~g, $10^{-8}$~g, $10^{-9}$~g,\,\ldots in such a way the maximal force $F_n$ for the mixed transition somewhat exceeds $F_{\rm min}$ (\ref{27.2b}). For the chosen mass $m$ we present the number of atoms $N$, sample activity $\alpha$, heat powers due to the electron-capture secondary products $W_e$ and $W_t$. The nuclei in the top and bottom parts of the Table are arranged in the ascending order of mass number.

One can readily see from Tables~\ref{t1}-\ref{t3} that the sample activity $\alpha$ varies but slightly: it is of the scale of $1$~MBq, even though $N$ may change by 5 orders of magnitude. This is because of Eq. (\ref{18}). Indeed, the sample activities $\alpha$ for different isotopes and transitions are to be comparable, if comparable are the recoil force and the values of $I_{n EC}$, $E_{\nu n}$, $B_n$, and $P$.

The situation is similar for the heat powers $W_e$ (\ref{25.2c}) and $W_t$ (\ref{25.2b}). Typically, the energies of secondary electrons are of the scale of some tens of keV, thus the corresponding heat load is of the scale of nW for $\alpha=1$~MBq (indeed, $1$~MBq $\cdot$ 10~keV $\simeq 1.6$~nW). It means that the radioactive decays contribution to the total heat load should not be a problem, because even for the temperature 25~mK the cooling power of modern dilution refrigerators is of the scale of tens $\mu$W \cite{DeAngelis2012}.

In principle, the recoil force can be increased by the use of strong intra-atomic magnetic fields. Their application was discussed in Ref.~\cite{DeAngelis2012}. There is information on hyperfine magnetic fields $22.5$~T for $^{57}$Co, $18.785$~T for $^{65}$Zn, $70.6$~T for $^{119}$Sb and $33.3$~T for $^{131}$Cs (see Table~1 from Ref.~\cite{DeAngelis2012}; the basic properties of these isotopes are also presented in Tables~\ref{t1}-\ref{t3} of the present work).

Note, however, that only for the isotope $^{57}$Co the hyperfine field arises in cobalt metal. In the other cases, the hyperfine magnetic field on isotopes of interest arises in the presence of other elements, in particular, Zn and Cs are to be in iron, Sb in a compound Pd$_2$MnSb. Therefore, in the case of Zn, Sb and Cs isotopes, radioactive atoms will account for only a fraction of the sample attached to a cantilever. At the same time, the advantage in the value of the magnetic field is not that significant compared to the field from superconducting magnets. Note also, that at $T=1$~K even for the $^{119}$Sb isotope in a hyperfine magnetic field $70.6$~T, the polarization reaches only $\sim 4$\,\%. Thus, the estimates reproduced above for completely polarized $^{119}$Sb nuclei are only valid if, following Ref.~\cite{DeAngelis2012}, one assumes that the temperature can be lowered to $25$~mK; in this case Eq. (\ref{20b}) leads to $P\simeq 87$\,\%.

The impact of intra-atomic fields can be easily evaluated by scaling the data presented in Tables~\ref{t1}-\ref{t3}. If, for instance, in a $^{57}$Co sample each nucleus is affected by the hyperfine field $B=22.5$~T (instead of 10~T) at the temperature $T=1$~K, then the sample with the mass $m=3.5\cdot 10^{-10}$~g (see Table~\ref{t2}), attached to a cantilever, will generate a recoil force $2.25\cdot 10^{-19}$~N (instead of $10^{-19}$~N).

\section{Potential applications}\label{s7}

\subsection{Neutrino mass. Fundamental symmetries}

For a pure Gamow--Teller transition, $J_i\to J_i\pm 1$, the asymmetry coefficient $B_n$ and the other factors determining the recoil force (\ref{18}) seem to be fully defined. So, in principle, a precise measurement of the recoil force for the isotopes from Tables~\ref{t1} and~\ref{t2} can give some information on the corrections to these factors. In reality, any factor is defined with some accuracy, thus the measurable magnitudes of corrections would be limited by the uncertainty of the recoil force.

Notice, in particular, that even the coefficient $B_n$ is not completely determined by Eq. (\ref{13}). Indeed, Eq. (\ref{13}) is obtained assuming the V-A variant and time reversal invariance (TRI) hold. In fact, tensor coupling as well as TRI violation may contribute to the coefficient $B_n$, as it seen from Ref. \cite{Treiman1958}. Current limitations for these factors are presented, e.g., in Ref. \cite{Vos2015b} (see also references therein). A consistent analysis of the accuracy of $B_n$ and all other factors entering Eq. (\ref{18}) is beyond the scope of this paper; we outlined the situation only to clarify the next issue related to the neutrino mass.

In principle, the recoil force depends on the neutrino mass via the neutrino momentum,
\begin{multline}\label{7b}
p_{\nu nx}=\sqrt{\left(\frac{E_{\nu nx}}{c}\right)^2 - (m_{\nu} c)^2} \\
\simeq \frac{E_{\nu n x}}{c}\left(1-\frac{1}{2}\left(\frac{m_{\nu}c^2}{E_{\nu nx}}\right)^2\right),
\end{multline}
that determines both the recoil momentum (\ref{17}) and the reduction factor $\eta_n$ (\ref{12.1}) for the asymmetry coefficient. To simplify, we take into account only $K$ capture and obtain
\begin{equation}\label{7c}
\eta_n=\frac{cp_{\nu nK}}{E_{\nu nK}}\simeq\,
1-\frac{1}{2}\left(\frac{m_{\nu}c^2}{E_{\nu nK}}\right)^2.
\end{equation}
We do not consider the impact of neutrino mass on the integral decay rate $w_{nEC}$ (due to phase space reduction) because we expressed it in Eq. (\ref{18}) in terms of the observed half-life $T_{1/2}$.

Thus, the recoil force includes a product of the factors (\ref{7b}) and (\ref{7c}) and have a somewhat smaller value for massive neutrino:
\begin{equation} \label{11b}
\frac{F_n(m_{\nu}\neq 0)}{F_n(m_{\nu}=0)}\simeq 1-\left(\frac{m_{\nu}c^2}{Q_{n EC}}\right)^2.
\end{equation}
The effect of the neutrino mass on the recoil force was discussed in \cite{Folan2014} (but in doing so the factor (\ref{7c}) was not considered) taking into account the existing limit $m_{\nu}c^2\le 2$~eV. Taking $Q_{n EC}\simeq 100$~keV, one obtains relative change in force of order of $\sim 10^{-10}$. Obviously, this value is much smaller than the accuracy with which the recoil force (\ref{18}) can be determined. For this reason, the neutrino mass is hardly unlikely to be  measured in this way.

However, unique contributions to the asymmetry of neutrino emission can be detected even if they are smaller than the uncertainty of the regular term $B_nP\, ({\bf n}_{\nu}{\bf n}_I)$ in~(\ref{12}), where ${\bf n}_{\nu}$ and ${\bf n}_I$ are unit vectors along neutrino momentum and nuclear polarization axis ($\eta_n$ is taken equal to one). Such contributions resulting from hypothetical Lorentz invariance violation were recently found in Ref.~\cite{Vos2015a}. One of them for a pure Gamow--Teller transition is of the form
\begin{equation}\label{7d}
B_n P\,\chi_i^{s0}\left[{\bf n}_{\nu}\times {\bf n}_I\right]_s,
\end{equation}
where $\chi_i^{s0}$ ($s=1,2,3$) are imaginary parts of the components of a complex tensor $\chi^{\mu\nu}$ which parametrizes Lorentz violation.

One can see from Fig.~\ref{f2} that measuring the recoil force for the polarizing magnetic field $B$ directed along the axis $y$ allows to detect or to set upper limit on the value of $\chi_i^{10}$ (according to Ref. \cite{Vos2015a}, at present this value is unconstrained). Clearly, the isotopes from Tables~\ref{t1} and~\ref{t2} (some of them were discussed in Ref. \cite{Vos2015a}) are the most suitable for such experiment.

\subsection{Probability distribution $P_x$}

The probabilities $P_x$ of electron capture from different shells $x$ are of specific interest because they are determined by purely atomic properties. In the approximation of point-like nucleus the ratio
\begin{equation}\label{39}
\frac{P_x}{P_{x^{\prime}}}= \frac{E_{\nu nx}^2}{E_{\nu nx^{\prime}}^2}\,\frac{|\psi_x(0)|^2}{|\psi_{x^{\prime}}(0)|^2}
\end{equation}
is sensitive to the electron wave functions on the nucleus. Such ratios are measured by detecting the secondary products of EC, i.e., Auger electrons and x-rays \cite{Bambynek1977}. Throughout the paper we used the evaluated probabilities $P_x$ (or the averaged energies $E_{\nu n}$ calculated with these evaluated probabilities) given on the website \cite{IAEA}. Of course, these evaluations are consistent with all accessible experimental data. However, an independent approach to determine the probabilities $P_x$ is of interest.

Measuring of the recoil force may be considered such an approach. Indeed, the force (\ref{18}) is proportional to the neutrino energy $E_{\nu n}$ (\ref{17.2}), averaged over $x$. Taking into account only the dominant contributions of $K$ and $L$ shells, one can obtain the probabilities $P_K$ and $P_L$ from the equations:
\begin{equation}\label{40}
\left\{\begin{array}{l}
P_KE_{\nu nK}+P_LE_{\nu nL}=E_{\nu n},\\
P_K+P_L=1.\\
\end{array}\right.
\end{equation}
Adding $P_M$ to $P_K$ and $P_L$ and assuming some relation between them, one can find the capture probabilities in such extended model.

Clearly, the larger is the difference between $E_{\nu nL}$ and $E_{\nu nK}$, the higher is the accuracy of this method. The difference increases with the charge of the radioactive nucleus and takes the values from 5.3 keV for the decay $^{54}$Mn\,$\to$\,$^{54}$Cr to 55.7 keV for the decay $^{179}$W\,$\to$\,$^{179}$Ta (we account for only pure Gamow--Teller decays included to Tables~\ref{t1} and~\ref{t2} each being determined by a definite asymmetry coefficient $B_n$).

\subsection{Mixing ratio for Fermi and Gamow--Teller contributions}

Isotopes from Table~\ref{t3} decay via mixed Fermi and Gamow--Teller transitions, which involves two reduced matrix elements. The decay rate is determined by the sum of the squared matrix elements, see Eq. (\ref{15}). By measuring the recoil force, one can obtain the asymmetry coefficient $B_n$ and, subsequently, the ratio of the reduced matrix elements (see Eq. (\ref{16})), taking the ratio $g_A/g_V=-1.27$ as known \cite{PDG2018}. Thus, Fermi and Gamow--Teller contributions to the decay rate can be distinguished. The two isotopes, $^{37}$Ar and $^{49}$V, from the top part of Table~\ref{t3} are the most suitables for such measurement.

In the bottom part of Table~\ref{t3}, three isotopes are presented that  undergo a pure Gamow--Teller transition (of type $J_i\to J_i-1$ for all three isotopes) or a mixed transition ($J_i\to J_i$). In this case, the recoil force is the sum of two contributions (\ref{18}), one of which corresponds to the Gamow-Teller transition and can be calculated explicitly. Therefore, the recoil force measurement probes, in fact, the coefficient $B_n$ (and the corresponding ratio of reduced nuclear matrix elements) for the mixed transition.

For the isotopes $^7$Be and $^{51}$Cr the Gamow--Teller transition occurs to an excited state of the daughter nucleus with a relatively small branching ratio, while the mixed transition occurs to the ground state of the daughter nucleus with a relatively high branching ratio. As a result, the force caused by the pure Gamow--Teller transition (which can be evaluated explicitly), is by one order of magnitude smaller than the maximum possible force caused by the mixed transition. For this reason, the situation is quite similar to the case of $^{37}$Ar and $^{49}$V.

As for the isotope $^{65}$Zn, both transitions have comparable branching ratios, but the pure Gamow--Teller transition occurs to the ground state of the daughter nucleus and produces much larger neutrino energy. Consequently, the force related to this transition (and explicitly calculable) turns out to be large ($F_0=46.5\cdot 10^{-19}$~N). In this case, in contrast to the previously discussed, a relatively small force (its maximal value $F_n=8.1\cdot 10^{-19}$~N is given in Table~\ref{t3}) caused by the mixed transition and dependent on the ratio of the reduced matrix elements (\ref{16}), is, in fact, a correction to $F_0$.

\section{Conclusion}\label{s8}

A sample of radioactive atoms experiences a recoil force from neutrino radiation accompanying electron capture by polarized nuclei provided there is a directional asymmetry of neutrino emission. This recoil is of interest for at least two reasons. First, the force is proportional to the asymmetry coefficient, i.e., the force measuring is equivalent to measuring of the neutrino angular distribution asymmetry. Second, the recoil is also proportional to neutrino momentum transfer to the sample. Thus, the force is sensitive to factors determining the momentum (the neutrino mass among them).

In this paper, we derive the asymmetry coefficient due to parity violation for allowed nuclear transitions taking into account the neutrino mass. An expression for the corresponding neutrino recoil force acting on the sample is obtained. We show that the magnitude of this force for pure Gamow--Teller transitions and for mixed Fermi and Gamow--Teller transitions are comparable. Prospects to measure this force by the use of modern micromechanical devices are discussed and appear to be realistic. Numerical estimates for the force are performed for a number of most suitable radioactive isotopes.

It is shown that, as it would be expected, the sensivity of the recoil force to neutrino mass is too small to detect the mass. However, another factor governing the neutrino momentum can be revealed. EC is, in fact, a transformation of the whole atom (not only the nucleus). Thus, the daughter atom gets one of the excited states each being related with a definite energy of the emitted neutrino. Since the recoil is determined by the neutrino energy averaged over the final atomic states, the force turns out to be a source of information on the probability distribution over these states.

Besides, as mentioned above, the recoil force is intimately related with the asymmetry of neutrino angular distribution. As for parity violation, the asymmetry coefficient for allowed nuclear transitions was first found in Ref. \cite{Treiman1958}. It has a definite value for a pure Gamow--Teller transition. However, for a mixed Fermi and Gamow--Teller transition, the coefficient is determined by the ratio of the corresponding reduced matrix elements. Thus for such a mixed transition, this ratio can be found by the recoil force measurement. In the same manner, the recoil force can give information on hypothetical Lorentz invariance violation resulting unique asymmetry terms in the neutrino angular distribution \cite{Vos2015a}.

We show that the recoil force being very small can be measured by the use of the methods developed in magnetic resonance force microscopy (MRFM). It means, in particular, that further improvements discussed in MRFM (see., e.g., \cite{Kuehn2008,Poggio2010}) could be used as well to increase accuracy of the neutrino recoil force measurements.

The first point to focus on is the nuclear polarization $P$. Throughout the paper we assumed that the polarization results from the Boltzmann distribution (\ref{19b}) at the temperature $T=1$~K in the magnetic field $B=10$~T. Under these conditions the polarization, as it is seen from Tables~\ref{t1} and~\ref{t2}, does not exceed $1\,\%$ (this also holds for the isotopes from Table~\ref{t3}). Meanwhile, the hyperfine interaction may be used to transfer the polarization from the electronic subsystem to the nuclear one. Different methods of such dynamical polarization, tested in MRFM experiments \cite{Thurber2003,Issac2016} at the temperature $\sim$~5~K, allowed to increase the nuclear polarization by the factor of~$\sim$~10. Since the recoil force (\ref{18}) is proportional to $P$, an enhancement of the polarization by 10-100 times by means of dynamical methods would increase the force by the same factor.

The second point to consider is reducing the threshold for the measured force (\ref{26b}). This requires temperature lowering, as well as increasing the quality factor~$Q$. To fulfill these requirements, one has to minimize the heat load by improving the system generating rf magnetic field, to develop new methods for cantilever displacement sensing and to probe new materials for microresonators production.

An additional factor can also contribute to the recoil force measurement improvement. MRFM is aimed at obtaining information about the magnetization distribution in a sample; this is only possible, if the magnetic force for every small part of the sample is measured quite quickly. Therefore, a bandwidth $\Delta\nu\simeq 1/\Delta t$ in (\ref{26b}) cannot be too narrow. The recoil force measurement, on the contrary, can be carried out for a long time, with an effective narrowing of $\Delta\nu$.

Thus, the measurement of the recoil force caused by the neutrino emission in EC by polarized nuclei is of interest for physics of the weak interaction as well as for the search of effects beyond the Standard Model. This approach can become a useful complement to the experiments of traditional type, since it employs methods fundamentally different from the ''standard'' ones used in nuclear and particle physics. In the case of successful realization, these methods may appear useful for the other studies dealing with small force measurements.

\begin{acknowledgments}
The work of O.A.T. was supported by the Russian Foundation for Basic Research under the project No. 18-32-00732.
\end{acknowledgments}

\bibliographystyle{apsrev} 
\bibliography{Refs_Neutrino_Recoil_v00}

\begin{thebibliography}{38}
\expandafter\ifx\csname natexlab\endcsname\relax\def\natexlab#1{#1}\fi
\expandafter\ifx\csname bibnamefont\endcsname\relax
  \def\bibnamefont#1{#1}\fi
\expandafter\ifx\csname bibfnamefont\endcsname\relax
  \def\bibfnamefont#1{#1}\fi
\expandafter\ifx\csname citenamefont\endcsname\relax
  \def\citenamefont#1{#1}\fi
\expandafter\ifx\csname url\endcsname\relax
  \def\url#1{\texttt{#1}}\fi
\expandafter\ifx\csname urlprefix\endcsname\relax\def\urlprefix{URL }\fi
\providecommand{\bibinfo}[2]{#2}
\providecommand{\eprint}[2][]{\url{#2}}

\bibitem[{\citenamefont{Bernabeu et~al.}(2005)\citenamefont{Bernabeu,
  Burguet-Castell, Espinoza, and Lindroos}}]{Bernabeu2005}
\bibinfo{author}{\bibfnamefont{J.}~\bibnamefont{Bernabeu}},
  \bibinfo{author}{\bibfnamefont{J.}~\bibnamefont{Burguet-Castell}},
  \bibinfo{author}{\bibfnamefont{C.}~\bibnamefont{Espinoza}}, \bibnamefont{and}
  \bibinfo{author}{\bibfnamefont{M.}~\bibnamefont{Lindroos}},
  \bibinfo{journal}{J. High Energy Phys.} \textbf{\bibinfo{volume}{2005}},
  \bibinfo{pages}{014} (\bibinfo{year}{2005}).

\bibitem[{\citenamefont{Sato}(2005)}]{Sato2005}
\bibinfo{author}{\bibfnamefont{J.}~\bibnamefont{Sato}}, \bibinfo{journal}{Phys.
  Rev. Lett.} \textbf{\bibinfo{volume}{95}}, \bibinfo{pages}{131804}
  (\bibinfo{year}{2005}).

\bibitem[{\citenamefont{Zucchelli}(2002)}]{Zucchelli2002}
\bibinfo{author}{\bibfnamefont{P.}~\bibnamefont{Zucchelli}},
  \bibinfo{journal}{Phys. Lett. B} \textbf{\bibinfo{volume}{532}},
  \bibinfo{pages}{166 } (\bibinfo{year}{2002}).

\bibitem[{\citenamefont{Edgecock et~al.}(2013)\citenamefont{Edgecock, Caretta,
  Davenne, Densam, Fitton, Kelliher, Loveridge, Machida, Prior, Rogers
  et~al.}}]{Edgecock2013}
\bibinfo{author}{\bibfnamefont{T.~R.} \bibnamefont{Edgecock}},
  \bibinfo{author}{\bibfnamefont{O.}~\bibnamefont{Caretta}},
  \bibinfo{author}{\bibfnamefont{T.}~\bibnamefont{Davenne}},
  \bibinfo{author}{\bibfnamefont{C.}~\bibnamefont{Densam}},
  \bibinfo{author}{\bibfnamefont{M.}~\bibnamefont{Fitton}},
  \bibinfo{author}{\bibfnamefont{D.}~\bibnamefont{Kelliher}},
  \bibinfo{author}{\bibfnamefont{P.}~\bibnamefont{Loveridge}},
  \bibinfo{author}{\bibfnamefont{S.}~\bibnamefont{Machida}},
  \bibinfo{author}{\bibfnamefont{C.}~\bibnamefont{Prior}},
  \bibinfo{author}{\bibfnamefont{C.}~\bibnamefont{Rogers}},
  \bibnamefont{et~al.}, \bibinfo{journal}{Phys. Rev. ST Accel. Beams}
  \textbf{\bibinfo{volume}{16}}, \bibinfo{pages}{021002}
  (\bibinfo{year}{2013}).

\bibitem[{\citenamefont{Wildner et~al.}(2014)\citenamefont{Wildner, Hansen,
  Benedetto, Jensen, Stora, Mendonca, Vlachoudis, Bouquerel, Marie-Jeanne,
  Balint et~al.}}]{Wildner2014}
\bibinfo{author}{\bibfnamefont{E.}~\bibnamefont{Wildner}},
  \bibinfo{author}{\bibfnamefont{C.}~\bibnamefont{Hansen}},
  \bibinfo{author}{\bibfnamefont{E.}~\bibnamefont{Benedetto}},
  \bibinfo{author}{\bibfnamefont{E.}~\bibnamefont{Jensen}},
  \bibinfo{author}{\bibfnamefont{T.}~\bibnamefont{Stora}},
  \bibinfo{author}{\bibfnamefont{T.~M.} \bibnamefont{Mendonca}},
  \bibinfo{author}{\bibfnamefont{V.}~\bibnamefont{Vlachoudis}},
  \bibinfo{author}{\bibfnamefont{E.}~\bibnamefont{Bouquerel}},
  \bibinfo{author}{\bibfnamefont{M.}~\bibnamefont{Marie-Jeanne}},
  \bibinfo{author}{\bibfnamefont{P.}~\bibnamefont{Balint}},
  \bibnamefont{et~al.}, \bibinfo{journal}{Phys. Rev. ST Accel. Beams}
  \textbf{\bibinfo{volume}{17}}, \bibinfo{pages}{071002}
  (\bibinfo{year}{2014}).

\bibitem[{\citenamefont{Barabanov and Titov}(2015)}]{Barabanov2015}
\bibinfo{author}{\bibfnamefont{A.~L.} \bibnamefont{Barabanov}}
  \bibnamefont{and} \bibinfo{author}{\bibfnamefont{O.~A.} \bibnamefont{Titov}},
  \bibinfo{journal}{Eur. Phys. J. A} \textbf{\bibinfo{volume}{51}},
  \bibinfo{pages}{96} (\bibinfo{year}{2015}).

\bibitem[{\citenamefont{Barabanov and Titov}(2017)}]{Barabanov2017}
\bibinfo{author}{\bibfnamefont{A.~L.} \bibnamefont{Barabanov}}
  \bibnamefont{and} \bibinfo{author}{\bibfnamefont{O.~A.} \bibnamefont{Titov}},
  \bibinfo{journal}{Phys. At. Nucl.} \textbf{\bibinfo{volume}{80}},
  \bibinfo{pages}{1181} (\bibinfo{year}{2017}).

\bibitem[{\citenamefont{DeAngelis et~al.}(2012)\citenamefont{DeAngelis, Folan,
  and Tsifrinovich}}]{DeAngelis2012}
\bibinfo{author}{\bibfnamefont{C.}~\bibnamefont{DeAngelis}},
  \bibinfo{author}{\bibfnamefont{L.~M.} \bibnamefont{Folan}}, \bibnamefont{and}
  \bibinfo{author}{\bibfnamefont{V.~I.} \bibnamefont{Tsifrinovich}},
  \bibinfo{journal}{Phys. Rev. C} \textbf{\bibinfo{volume}{86}},
  \bibinfo{pages}{034615} (\bibinfo{year}{2012}).

\bibitem[{\citenamefont{Folan and Tsifrinovich}(2014)}]{Folan2014}
\bibinfo{author}{\bibfnamefont{L.~M.} \bibnamefont{Folan}} \bibnamefont{and}
  \bibinfo{author}{\bibfnamefont{V.~I.} \bibnamefont{Tsifrinovich}},
  \bibinfo{journal}{Mod. Phys. Lett. A} \textbf{\bibinfo{volume}{29}},
  \bibinfo{pages}{1430042} (\bibinfo{year}{2014}).

\bibitem[{\citenamefont{Folan and Tsifrinovich}(2017)}]{Folan2017}
\bibinfo{author}{\bibfnamefont{L.~M.} \bibnamefont{Folan}} \bibnamefont{and}
  \bibinfo{author}{\bibfnamefont{V.~I.} \bibnamefont{Tsifrinovich}},
  \bibinfo{journal}{World Journal of Nuclear Science and Technology}
  \textbf{\bibinfo{volume}{7}}, \bibinfo{pages}{58} (\bibinfo{year}{2017}).

\bibitem[{\citenamefont{Allen}(1942)}]{Allen1942}
\bibinfo{author}{\bibfnamefont{J.~S.} \bibnamefont{Allen}},
  \bibinfo{journal}{Phys. Rev.} \textbf{\bibinfo{volume}{61}},
  \bibinfo{pages}{692} (\bibinfo{year}{1942}).

\bibitem[{\citenamefont{Englert et~al.}(2018)\citenamefont{Englert, Hild, and
  Spannowsky}}]{Englert2018}
\bibinfo{author}{\bibfnamefont{C.}~\bibnamefont{Englert}},
  \bibinfo{author}{\bibfnamefont{S.}~\bibnamefont{Hild}}, \bibnamefont{and}
  \bibinfo{author}{\bibfnamefont{M.}~\bibnamefont{Spannowsky}},
  \bibinfo{journal}{EPL} \textbf{\bibinfo{volume}{123}}, \bibinfo{pages}{41001}
  (\bibinfo{year}{2018}).

\bibitem[{\citenamefont{Treiman}(1958)}]{Treiman1958}
\bibinfo{author}{\bibfnamefont{S.~B.} \bibnamefont{Treiman}},
  \bibinfo{journal}{Phys. Rev.} \textbf{\bibinfo{volume}{110}},
  \bibinfo{pages}{448} (\bibinfo{year}{1958}).

\bibitem[{\citenamefont{Vos et~al.}(2015{\natexlab{a}})\citenamefont{Vos,
  Wilschut, and Timmermans}}]{Vos2015a}
\bibinfo{author}{\bibfnamefont{K.~K.} \bibnamefont{Vos}},
  \bibinfo{author}{\bibfnamefont{H.~W.} \bibnamefont{Wilschut}},
  \bibnamefont{and} \bibinfo{author}{\bibfnamefont{R.~G.~E.}
  \bibnamefont{Timmermans}}, \bibinfo{journal}{Phys. Rev. C}
  \textbf{\bibinfo{volume}{91}}, \bibinfo{pages}{038501}
  (\bibinfo{year}{2015}{\natexlab{a}}).

\bibitem[{\citenamefont{Eisenberg and Greiner}(1976)}]{Eisenberg1976}
\bibinfo{author}{\bibfnamefont{J.}~\bibnamefont{Eisenberg}} \bibnamefont{and}
  \bibinfo{author}{\bibfnamefont{W.}~\bibnamefont{Greiner}},
  \emph{\bibinfo{title}{Nuclear Theory. Vol.~2. Excitation Mechanisms of the
  Nucleus}} (\bibinfo{publisher}{North-Holland Publishing Company},
  \bibinfo{address}{Amsterdam}, \bibinfo{year}{1976}).

\bibitem[{\citenamefont{Barabanov et~al.}(1996)\citenamefont{Barabanov,
  Gaponov, Danilin, and Shul'gina}}]{Barabanov1996}
\bibinfo{author}{\bibfnamefont{A.~L.} \bibnamefont{Barabanov}},
  \bibinfo{author}{\bibfnamefont{{\relax Yu}.~V.} \bibnamefont{Gaponov}},
  \bibinfo{author}{\bibfnamefont{B.~V.} \bibnamefont{Danilin}},
  \bibnamefont{and} \bibinfo{author}{\bibfnamefont{N.~B.}
  \bibnamefont{Shul'gina}}, \bibinfo{journal}{Phys. At. Nucl.}
  \textbf{\bibinfo{volume}{59}}, \bibinfo{pages}{1871} (\bibinfo{year}{1996}).

\bibitem[{\citenamefont{Barabanov}(2000)}]{Barabanov2000}
\bibinfo{author}{\bibfnamefont{A.~L.} \bibnamefont{Barabanov}},
  \bibinfo{journal}{Phys. At. Nucl.} \textbf{\bibinfo{volume}{63}},
  \bibinfo{pages}{1187} (\bibinfo{year}{2000}).

\bibitem[{\citenamefont{Bohr and Mottelson}(1998)}]{Bohr1998}
\bibinfo{author}{\bibfnamefont{A.}~\bibnamefont{Bohr}} \bibnamefont{and}
  \bibinfo{author}{\bibfnamefont{B.~R.} \bibnamefont{Mottelson}},
  \emph{\bibinfo{title}{Nuclear Structure. Vol.~1. Single-Particle Motion.}}
  (\bibinfo{publisher}{World Scientific}, \bibinfo{year}{1998}).

\bibitem[{\citenamefont{Kittel}(2005)}]{Kittel2005}
\bibinfo{author}{\bibfnamefont{C.}~\bibnamefont{Kittel}},
  \emph{\bibinfo{title}{Introduction to solid state physics}}
  (\bibinfo{publisher}{John Wiley \& Sons}, \bibinfo{year}{2005}).

\bibitem[{IAE()}]{IAEA}
\emph{\bibinfo{title}{Live chart of nuclides}},
  \bibinfo{howpublished}{https://www-nds.iaea.org}.

\bibitem[{\citenamefont{Bearden and Burr}(1967)}]{Bearden1967}
\bibinfo{author}{\bibfnamefont{J.~A.} \bibnamefont{Bearden}} \bibnamefont{and}
  \bibinfo{author}{\bibfnamefont{A.~F.} \bibnamefont{Burr}},
  \bibinfo{journal}{Rev. Mod. Phys.} \textbf{\bibinfo{volume}{39}},
  \bibinfo{pages}{125} (\bibinfo{year}{1967}).

\bibitem[{\citenamefont{Hammel and Pelekhov}(2007)}]{Hammel2007}
\bibinfo{author}{\bibfnamefont{P.~C.} \bibnamefont{Hammel}} \bibnamefont{and}
  \bibinfo{author}{\bibfnamefont{D.~V.} \bibnamefont{Pelekhov}}, in
  \emph{\bibinfo{booktitle}{Handbook of Magnetism and Advanced Magnetic
  Materials}}, edited by
  \bibinfo{editor}{\bibfnamefont{H.}~\bibnamefont{Kronmuller}},
  \bibinfo{editor}{\bibfnamefont{S.}~\bibnamefont{Parkin}},
  \bibinfo{editor}{\bibfnamefont{S.}~\bibnamefont{Parkin}}, \bibnamefont{and}
  \bibinfo{editor}{\bibfnamefont{D.~D.} \bibnamefont{Awschalom}}
  (\bibinfo{publisher}{John Wiley \& Sons}, \bibinfo{year}{2007}), vol.
  \bibinfo{volume}{5: Spintronics and Magnetoelectronics}, \bibinfo{note}{part
  4: Quantum computation. The Magnetic Resonance Force Microscope}.

\bibitem[{\citenamefont{Ohnesorge and Binnig}(1993)}]{Ohnesorge1993}
\bibinfo{author}{\bibfnamefont{F.}~\bibnamefont{Ohnesorge}} \bibnamefont{and}
  \bibinfo{author}{\bibfnamefont{G.}~\bibnamefont{Binnig}},
  \bibinfo{journal}{Science} \textbf{\bibinfo{volume}{260}},
  \bibinfo{pages}{1451} (\bibinfo{year}{1993}).

\bibitem[{\citenamefont{Sidles et~al.}(1995)\citenamefont{Sidles, Garbini,
  Bruland, Rugar, Z\"uger, Hoen, and Yannoni}}]{Sidles1995}
\bibinfo{author}{\bibfnamefont{J.~A.} \bibnamefont{Sidles}},
  \bibinfo{author}{\bibfnamefont{J.~L.} \bibnamefont{Garbini}},
  \bibinfo{author}{\bibfnamefont{K.~J.} \bibnamefont{Bruland}},
  \bibinfo{author}{\bibfnamefont{D.}~\bibnamefont{Rugar}},
  \bibinfo{author}{\bibfnamefont{O.}~\bibnamefont{Z\"uger}},
  \bibinfo{author}{\bibfnamefont{S.}~\bibnamefont{Hoen}}, \bibnamefont{and}
  \bibinfo{author}{\bibfnamefont{C.~S.} \bibnamefont{Yannoni}},
  \bibinfo{journal}{Rev. Mod. Phys.} \textbf{\bibinfo{volume}{67}},
  \bibinfo{pages}{249} (\bibinfo{year}{1995}).

\bibitem[{\citenamefont{Suter}(2004)}]{Suter2004}
\bibinfo{author}{\bibfnamefont{A.}~\bibnamefont{Suter}},
  \bibinfo{journal}{Progress in Nuclear Magnetic Resonance Spectroscopy}
  \textbf{\bibinfo{volume}{45}}, \bibinfo{pages}{239 } (\bibinfo{year}{2004}).

\bibitem[{\citenamefont{Greenberg et~al.}(2012)\citenamefont{Greenberg,
  Pashkin, and Il'ichev}}]{Greenberg2012}
\bibinfo{author}{\bibfnamefont{{\relax Ya}.}~\bibnamefont{Greenberg}},
  \bibinfo{author}{\bibfnamefont{{\relax Yu}.~A.} \bibnamefont{Pashkin}},
  \bibnamefont{and} \bibinfo{author}{\bibfnamefont{E.}~\bibnamefont{Il'ichev}},
  \bibinfo{journal}{Phys. Usp.} \textbf{\bibinfo{volume}{55}},
  \bibinfo{pages}{382} (\bibinfo{year}{2012}).

\bibitem[{\citenamefont{Stowe et~al.}(1997)\citenamefont{Stowe, Yasumura,
  Kenny, Botkin, Wago, and Rugar}}]{Stowe1997}
\bibinfo{author}{\bibfnamefont{T.~D.} \bibnamefont{Stowe}},
  \bibinfo{author}{\bibfnamefont{K.}~\bibnamefont{Yasumura}},
  \bibinfo{author}{\bibfnamefont{T.~W.} \bibnamefont{Kenny}},
  \bibinfo{author}{\bibfnamefont{D.}~\bibnamefont{Botkin}},
  \bibinfo{author}{\bibfnamefont{K.}~\bibnamefont{Wago}}, \bibnamefont{and}
  \bibinfo{author}{\bibfnamefont{D.}~\bibnamefont{Rugar}},
  \bibinfo{journal}{Appl. Phys. Lett.} \textbf{\bibinfo{volume}{71}},
  \bibinfo{pages}{288} (\bibinfo{year}{1997}).

\bibitem[{\citenamefont{Mozyrsky et~al.}(2003)\citenamefont{Mozyrsky, Martin,
  Pelekhov, and Hammel}}]{Mozyrsky2003}
\bibinfo{author}{\bibfnamefont{D.}~\bibnamefont{Mozyrsky}},
  \bibinfo{author}{\bibfnamefont{I.}~\bibnamefont{Martin}},
  \bibinfo{author}{\bibfnamefont{D.}~\bibnamefont{Pelekhov}}, \bibnamefont{and}
  \bibinfo{author}{\bibfnamefont{P.~C.} \bibnamefont{Hammel}},
  \bibinfo{journal}{Appl. Phys. Lett.} \textbf{\bibinfo{volume}{82}},
  \bibinfo{pages}{1278} (\bibinfo{year}{2003}).

\bibitem[{\citenamefont{Mamin et~al.}(2005)\citenamefont{Mamin, Budakian, Chui,
  and Rugar}}]{Mamin2005}
\bibinfo{author}{\bibfnamefont{H.~J.} \bibnamefont{Mamin}},
  \bibinfo{author}{\bibfnamefont{R.}~\bibnamefont{Budakian}},
  \bibinfo{author}{\bibfnamefont{B.~W.} \bibnamefont{Chui}}, \bibnamefont{and}
  \bibinfo{author}{\bibfnamefont{D.}~\bibnamefont{Rugar}},
  \bibinfo{journal}{Phys. Rev. B} \textbf{\bibinfo{volume}{72}},
  \bibinfo{pages}{024413} (\bibinfo{year}{2005}).

\bibitem[{\citenamefont{Mamin et~al.}(2007)\citenamefont{Mamin, Poggio, Degen,
  and Rugar}}]{Mamin2007}
\bibinfo{author}{\bibfnamefont{H.~J.} \bibnamefont{Mamin}},
  \bibinfo{author}{\bibfnamefont{M.}~\bibnamefont{Poggio}},
  \bibinfo{author}{\bibfnamefont{C.~L.} \bibnamefont{Degen}}, \bibnamefont{and}
  \bibinfo{author}{\bibfnamefont{D.}~\bibnamefont{Rugar}},
  \bibinfo{journal}{Nature Nanotechnology} \textbf{\bibinfo{volume}{2}},
  \bibinfo{pages}{301} (\bibinfo{year}{2007}).

\bibitem[{\citenamefont{Xue et~al.}(2011)\citenamefont{Xue, Weber, Peddibhotla,
  and Poggio}}]{Xue2011}
\bibinfo{author}{\bibfnamefont{F.}~\bibnamefont{Xue}},
  \bibinfo{author}{\bibfnamefont{D.~P.} \bibnamefont{Weber}},
  \bibinfo{author}{\bibfnamefont{P.}~\bibnamefont{Peddibhotla}},
  \bibnamefont{and} \bibinfo{author}{\bibfnamefont{M.}~\bibnamefont{Poggio}},
  \bibinfo{journal}{Phys. Rev. B} \textbf{\bibinfo{volume}{84}},
  \bibinfo{pages}{205328} (\bibinfo{year}{2011}).

\bibitem[{\citenamefont{Thurber et~al.}(2003)\citenamefont{Thurber, Harrell,
  and Smith}}]{Thurber2003}
\bibinfo{author}{\bibfnamefont{K.~R.} \bibnamefont{Thurber}},
  \bibinfo{author}{\bibfnamefont{L.~E.} \bibnamefont{Harrell}},
  \bibnamefont{and} \bibinfo{author}{\bibfnamefont{D.~D.} \bibnamefont{Smith}},
  \bibinfo{journal}{Journal of Magnetic Resonance}
  \textbf{\bibinfo{volume}{162}}, \bibinfo{pages}{336 } (\bibinfo{year}{2003}).

\bibitem[{\citenamefont{Vos et~al.}(2015{\natexlab{b}})\citenamefont{Vos,
  Wilschut, and Timmermans}}]{Vos2015b}
\bibinfo{author}{\bibfnamefont{K.}~\bibnamefont{Vos}},
  \bibinfo{author}{\bibfnamefont{H.}~\bibnamefont{Wilschut}}, \bibnamefont{and}
  \bibinfo{author}{\bibfnamefont{R.}~\bibnamefont{Timmermans}},
  \bibinfo{journal}{Rev. Mod. Phys.} \textbf{\bibinfo{volume}{87}},
  \bibinfo{pages}{1483} (\bibinfo{year}{2015}{\natexlab{b}}).

\bibitem[{\citenamefont{Bambynek et~al.}(1977)\citenamefont{Bambynek, Behrens,
  Chen, Crasemann, Fitzpatrick, Ledingham, Genz, Mutterer, and
  Intemann}}]{Bambynek1977}
\bibinfo{author}{\bibfnamefont{W.}~\bibnamefont{Bambynek}},
  \bibinfo{author}{\bibfnamefont{H.}~\bibnamefont{Behrens}},
  \bibinfo{author}{\bibfnamefont{M.~H.} \bibnamefont{Chen}},
  \bibinfo{author}{\bibfnamefont{B.}~\bibnamefont{Crasemann}},
  \bibinfo{author}{\bibfnamefont{M.~L.} \bibnamefont{Fitzpatrick}},
  \bibinfo{author}{\bibfnamefont{K.~W.~D.} \bibnamefont{Ledingham}},
  \bibinfo{author}{\bibfnamefont{H.}~\bibnamefont{Genz}},
  \bibinfo{author}{\bibfnamefont{M.}~\bibnamefont{Mutterer}}, \bibnamefont{and}
  \bibinfo{author}{\bibfnamefont{R.~L.} \bibnamefont{Intemann}},
  \bibinfo{journal}{Rev. Mod. Phys.} \textbf{\bibinfo{volume}{49}},
  \bibinfo{pages}{77} (\bibinfo{year}{1977}).

\bibitem[{\citenamefont{Tanabashi et~al.}(2018)\citenamefont{Tanabashi,
  Hagiwara, Hikasa, Nakamura, Sumino, Takahashi, Tanaka, Agashe, Aielli, Amsler
  et~al.}}]{PDG2018}
\bibinfo{author}{\bibfnamefont{M.}~\bibnamefont{Tanabashi}},
  \bibinfo{author}{\bibfnamefont{K.}~\bibnamefont{Hagiwara}},
  \bibinfo{author}{\bibfnamefont{K.}~\bibnamefont{Hikasa}},
  \bibinfo{author}{\bibfnamefont{K.}~\bibnamefont{Nakamura}},
  \bibinfo{author}{\bibfnamefont{Y.}~\bibnamefont{Sumino}},
  \bibinfo{author}{\bibfnamefont{F.}~\bibnamefont{Takahashi}},
  \bibinfo{author}{\bibfnamefont{J.}~\bibnamefont{Tanaka}},
  \bibinfo{author}{\bibfnamefont{K.}~\bibnamefont{Agashe}},
  \bibinfo{author}{\bibfnamefont{G.}~\bibnamefont{Aielli}},
  \bibinfo{author}{\bibfnamefont{C.}~\bibnamefont{Amsler}},
  \bibnamefont{et~al.} (\bibinfo{collaboration}{Particle Data Group}),
  \bibinfo{journal}{Phys. Rev. D} \textbf{\bibinfo{volume}{98}},
  \bibinfo{pages}{030001} (\bibinfo{year}{2018}).

\bibitem[{\citenamefont{Kuehn et~al.}(2008)\citenamefont{Kuehn, Hickman, and
  Marohn}}]{Kuehn2008}
\bibinfo{author}{\bibfnamefont{S.}~\bibnamefont{Kuehn}},
  \bibinfo{author}{\bibfnamefont{S.~A.} \bibnamefont{Hickman}},
  \bibnamefont{and} \bibinfo{author}{\bibfnamefont{J.~A.}
  \bibnamefont{Marohn}}, \bibinfo{journal}{J. Chem. Phys.}
  \textbf{\bibinfo{volume}{128}}, \bibinfo{pages}{052208}
  (\bibinfo{year}{2008}).

\bibitem[{\citenamefont{Poggio and Degen}(2010)}]{Poggio2010}
\bibinfo{author}{\bibfnamefont{M.}~\bibnamefont{Poggio}} \bibnamefont{and}
  \bibinfo{author}{\bibfnamefont{C.~L.} \bibnamefont{Degen}},
  \bibinfo{journal}{Nanotechnology} \textbf{\bibinfo{volume}{21}},
  \bibinfo{pages}{342001} (\bibinfo{year}{2010}).

\bibitem[{\citenamefont{Issac et~al.}(2016)\citenamefont{Issac, Gleave, Nasr,
  Nguyen, Curley, Yoder, Moore, Chen, and Marohn}}]{Issac2016}
\bibinfo{author}{\bibfnamefont{C.~E.} \bibnamefont{Issac}},
  \bibinfo{author}{\bibfnamefont{C.~M.} \bibnamefont{Gleave}},
  \bibinfo{author}{\bibfnamefont{P.~T.} \bibnamefont{Nasr}},
  \bibinfo{author}{\bibfnamefont{H.~L.} \bibnamefont{Nguyen}},
  \bibinfo{author}{\bibfnamefont{E.~A.} \bibnamefont{Curley}},
  \bibinfo{author}{\bibfnamefont{J.~L.} \bibnamefont{Yoder}},
  \bibinfo{author}{\bibfnamefont{E.~W.} \bibnamefont{Moore}},
  \bibinfo{author}{\bibfnamefont{L.}~\bibnamefont{Chen}}, \bibnamefont{and}
  \bibinfo{author}{\bibfnamefont{J.~A.} \bibnamefont{Marohn}},
  \bibinfo{journal}{Phys. Chem. Chem. Phys.} \textbf{\bibinfo{volume}{18}},
  \bibinfo{pages}{8806} (\bibinfo{year}{2016}).

\end{thebibliography}

\end{document}